\documentclass[12pt]{iopart}
\usepackage{graphicx}
\usepackage{dcolumn}
\usepackage{bm}
\usepackage{color}
\usepackage{url}
\usepackage{hyperref}
\usepackage{array}

\expandafter\let\csname equation*\endcsname=\relax
\expandafter\let\csname endequation*\endcsname=\relax
\usepackage{amsmath,amssymb,amsthm} 
\usepackage{braket}
\usepackage{siunitx}

\begin{document}

\title[Improvements of PQCG]{Improvements of the programmable quantum current generator for better traceability of electrical current measurements}
\author{Sophie Djordjevic$^{1}$, Ralf Behr$^{2}$, Dietmar Drung$^{3}$, Martin G\"{o}tz$^{2}$ and Wilfrid Poirier$^{1}$}

\address{(1) Laboratoire de m\'{e}trologie et d'essais (LNE), 29 avenue Roger Hennequin, 78197 Trappes, France}
\address{(2) Physikalisch-Technische Bundesanstalt (PTB), Bundesallee 100, 38116 Braunschweig, Germany}
\address{(3) Physikalisch-Technische Bundesanstalt (PTB), Abbestra{\ss}e 2-12, 10587 Berlin, Germany}
\ead{sophie.djordjevic@lne.fr}
\vspace{10pt}
\begin{indented}
\item[]February 2021
\end{indented}

\begin{abstract}
A programmable quantum current generator based on the application of Ohm's law to quantum voltage and resistance standards has demonstrated a realization of the ampere from the elementary charge with a $10^{-8}$ relative uncertainty \cite{Brun-Picard2016}. Here, we report on improvements of the device leading to a noise reduction of the generated quantized current. The improved quantum current generator is used to calibrate different ammeters with lower measurement uncertainties. Besides, measurements of its quantized current using a calibrated Ultrastable Low-Noise Current Amplifier (ULCA) have shown that the realizations of the ampere at PTB (Physikalisch-Technische Bundesanstalt) and LNE (Laboratoire national de m\'{e}trologie et d'essais) in the range $\pm50$ $\mu$A agreed to -3.7 parts in $10^{7}$ with a combined standard uncertainty of 3.1 parts in $10^{7}$ (coverage factor $k_\mathrm{c}=1$).
\end{abstract}


%

\section{Introduction}

Since 20 May 2019, the ampere definition has been related to the quantum description of the electrical transport and not to a mechanical force anymore. As the numerical value of the elementary charge $e$ has been fixed, a current of one ampere is the flow of $1/1.602176634\times10^{-19}$ elementary charges per second \cite{SI_Brochure_2019}. This new definition has a big impact on the realization of the ampere \cite{Poirier2019} which can be based on solid state quantum effects, and in particular on application of Ohm's law to the Josephson voltage standard and the quantum Hall resistance standard, which are realizations of the volt and the ohm, directly related to defining constants of the SI : the Planck constant $h$ and the elementary charge $e$.
\begin{figure*}[!h]
  \centering
  \includegraphics[width=6.2in]{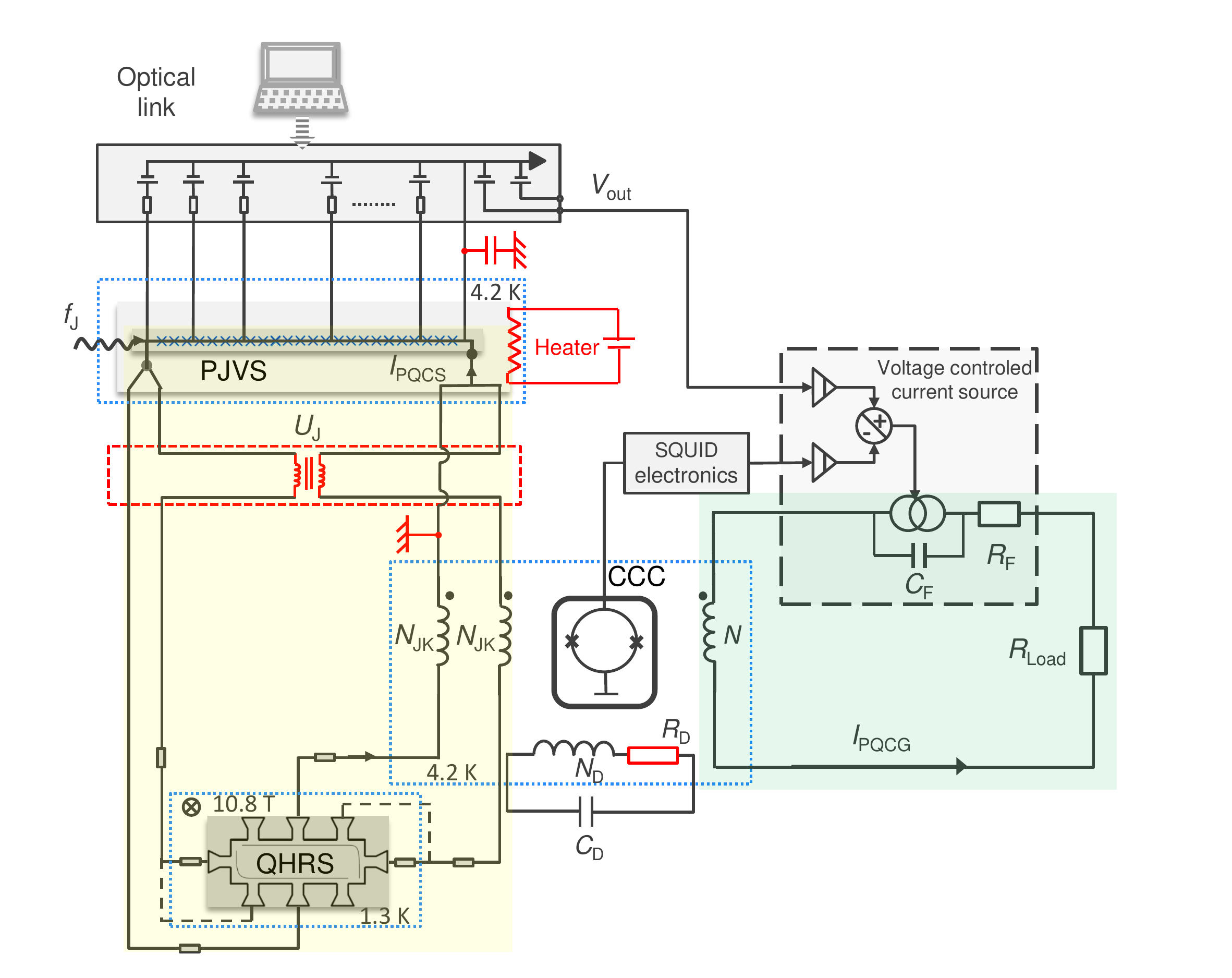}
  \caption{\small{Schematic of the PQCG with recent modifications (highlighted in red) detailed in section \ref{Improvements}. Note that $R_\mathrm{D}$ has not been introduced as a new element in the damping circuit but placed at low temperature rather than at ambient temperature. The primary quantum circuit PQCS and the secondary circuit are highlighted by yellow and green boxes, respectively.}}
  \label{fig:schema}
\end{figure*}
Implementing experimentally Ohm's law to quantum electrical standards, we have demonstrated that currents could be quantized in terms of $e$ within a relative uncertainty of $10^{-8}$ \cite{Brun-Picard2016}. This 'mise en pratique' of the ampere combines a programmable Josephson voltage standard (PJVS), a quantum Hall resistance standard (QHRS) and a cryogenic current comparator (CCC). The high accuracy of the generated current relies not only on the universality of the Josephson effect and the quantum Hall effect but also on the CCC accuracy itself. The latter is used as a highly accurate current amplifier with a gain span of more than four orders of magnitude. Quantized currents from mA to $\mu$A levels can therefore be generated with the same expected accuracy.

For small currents, below 1 nA, single electron current sources offer another way for the quantum realization of the definition of the ampere by controlling the flow of electrons one by one at a clock frequency (for recent reviews \cite{Pekola2013,Scherer2019,Giblin_2019,Kaneko_2016}). However, their principle of operation is based on the single electron tunneling which is a stochastic process, making the control of the electron flow difficult. Hence achieving an accuracy better than $10^{-7}$ remains very challenging \cite{Giblin_2019, Stein2017}.

For currents above 1 nA, more recently different groups also proposed developments based on Ohm's law involving quantum electrical standards, either to improve current measurement uncertainty in the $\mu$A range below $10^{-7}$ by current-to-voltage conversion with a quantum Hall resistance array \cite{Chae_2020} or to improve the stability of an externally referenced current source down to 1 nA/A at 50 mA with a PJVS \cite{Fan_2019}. On the other hand, National Metrology Institutes apply Ohm's law to secondary voltage and resistance standards, traceable to the Josephson and quantum Hall effects, respectively. However, the uncertainties claimed in their calibration and measurement capabilities (CMC) for currents are about $10^{-6}$. In this context, the recent developments of the ULCA at PTB \cite{DrungRSI2015} based on Ohm's law tackle this limit for current levels up to 50 $\mu$A. The first version of the amplifier was specially focused on small currents $\leq$ 5 nA, since it was dedicated to amplify and measure the very small current generated by single-electron pumps, but recent developments increased the range up to 50 $\mu$A \cite{Drung2017}. The predicted relative uncertainties for the ULCA devices from the nA range to the $\mu$A range is 2 parts in $10^7$ for timely calibrations ($k_\mathrm{c}=1$). All these developments have opened the way to new current comparisons targeting measurement uncertainties of few parts in $10^{7}$.

The first demonstration of the PQCG showed very promising results, however, making it a primary current standard needs to ensure a very reliable operation. The reduction of noise and instability are essential for that purpose. At first, the paper details the noise characterizations and the modifications implemented to improve the set-up. It then reports on ammeter calibrations carried out following well established quantization criteria, that notably demonstrate an improvement of the measurement uncertainties in the mA range. Finally, a first direct comparison of the ampere realizations performed with measurement uncertainties of a few parts in $10^{7}$ at 50 $\mu$A is presented.

\section{Improving the stability and reliability of the set-up}

\subsection{The set-up}
The principle of the programmable quantum current generator is presented in figure\space\ref{fig:schema}. It is based on a primary circuit, the programmable quantum current standard (PQCS), highlighted by the light yellow box, where the quantum reference current $I_{\mathrm{PQCS}}$ is circulating. This circuit is magnetically linked by the cryogenic current amplifier to a second circuit that drives the output current of the PQCG, $I_{\mathrm{PQCG}}$. In the reference loop, at the superconducting pads of the PJVS, quantized voltage steps, $U_\mathrm{J} = n_\mathrm{J}f_\mathrm{J}K^{-1}_\mathrm{J}$ with $K_\mathrm{J} = (2e/h)$, are generated when $n_\mathrm{J}$ Josephson junctions are dc biased and driven at a microwave frequency $f_\mathrm{J}$. This voltage is applied to the QHRS through a double connection scheme \cite{Delahaye1993,Poirier2014}. Doing so, and thanks to the properties of the quantum Hall effect \cite{Poirier2014,Delahaye1993}, the equivalent resistance $R$ connected to the superconducting pads is equal to $R = R_\mathrm{H}(1+\alpha)$, where $R_\mathrm{H}$ is the quantum Hall resistance, which is equal to $R_\mathrm{K}/2$ at Landau level filling factor $\nu=2$ with $R_\mathrm{K} = h/e^2$ the von Klitzing constant, and where $\alpha=\mathcal{O}((r/R_\mathrm{H})^2)$ is a second order correction due to the cable resistances $r$. This correction, which is significatively reduced compared to the usual first order correction of a simple connection corresponding to $r/R_\mathrm{H} \sim 3\times 10^{-4}$, can be determined with a relative uncertainty of a few parts in $10^{9}$.

The reference current $I_\mathrm{PQCS}=\frac{U_\mathrm{J}}{R} = \frac{U_\mathrm{J}}{R_\mathrm{H}}(1-\alpha)$ is circulating into the PJVS. It can also be written as $I_\mathrm{PQCS} = n_\mathrm{J} e f_\mathrm{J} (1-\alpha)$, which emphasizes the relation to the elementary charge. This current is divided into two separate branches in the double connection to the QHRS. In order to detect and amplify the sum of these currents, we use a CCC, which insures the ampere-turns balance of several superconducting windings by means of the SQUID-based feedback loop. Thanks to the introduction of two CCC windings of identical number of turns, $N_\mathrm{JK}$, into the double connection between the PJVS and the QHRS, the balance between the primary and secondary circuit, $N_\mathrm{JK} I_\mathrm{PQCS} - N I_\mathrm{PQCG} = 0$, leads to $I_\mathrm{PQCG} = \frac{N_\mathrm{JK}}{N} I_\mathrm{PQCS}$. The gain $G = \frac{N_\mathrm{JK}}{N}$ is exact within one part in $10^{9}$ for all windings and within few parts in $10^{10}$ for the windings used in the present study \cite{Poirier2020}. The current $I_\mathrm{PQCG}$ is generated by an external battery-powered low-noise current source which is servo-controlled by the feedback voltage of the SQUID that monitors the flux imbalance. An adjusted pre-balance current (or compensation current), $I_\mathrm{comp}$, driven by the voltage $V_\mathrm{out}$ supplied by a digital-to-analog converter (DAC) of the programmable current bias source of the PJVS, is added to the feedback current of the SQUID to reduce the set-point error due to the finite open loop gain of the SQUID electronics \cite{Brun-Picard2016}. The current $I_\mathrm{PQCG}$ circulates in the load which can be a resistance $R_\mathrm{Load}$, an ammeter or the ULCA.

The PJVS used in this work is the PJVS$\sharp$A from reference \cite{Brun-Picard2016}. It is based on 1 V Nb/Al/AlOx/Al/AlOx/Al/Nb Josephson junction series arrays fabricated at the PTB \cite{Mueller2007,Muller2009}. The QHRS is described in reference \cite{Piquemal1993} and the CCC details can be found in \cite{Poirier2020}. More details are given in section \ref{exp_details}.

In our previous work \cite{Brun-Picard2016}, the correct operation of the PQCG was achieved by solving several technical issues. As the PJVS, the QHRS and the CCC are implemented in three different cryostats and connected by long shielded cables, it makes the PQCG very sensitive to external electromagnetic noise. The stability issues are more severe for high ampere-turns values. For instance, the SQUID became unstable for $N_\mathrm{JK}>2$ when the PQCS circuit was connected. However, increasing $N_\mathrm{JK}$ is relevant to improve the signal-to-noise ratio of the PQCG, which is given by the ratio of the flux generated into the SQUID by the primary circuit, $N_{\mathrm{JK}}I_{\mathrm{PQCS}}/\gamma_\mathrm{CCC}$ ($\gamma_\mathrm{CCC}$ being the flux to ampere-turn sensitivity of the CCC), to the total flux noise $\sqrt{S^\mathrm{tot}_{\phi}(f)}$. To solve this problem, a damping circuit was implemented that stabilizes the CCC operation. Since it can not be added directly on the reference circuit without introducing systematic errors, a series arrangement of a resistance $R_\mathrm{D}$  and a capacitance $C_\mathrm{D}$ has been connected to a spare winding of $N_\mathrm{D}$ turns \cite{Drung2009} as shown in figure\space\ref{fig:schema}. However, the damping circuit introduced increased noise in the 100 Hz - 10 kHz bandwidth, the consequence of which was not analysed in detailed at that time. Besides, it remained impossible to increase $N_\mathrm{JK}$ above 128. Moreover, the low-frequency noise was significantly higher than the SQUID specifications. Finally, the repeatability of the PQCG turned out to be restricted due to flux frequently trapped in the Josephson array during the measurement campaigns. In the following, we report on investigations about noise, stability issues and resonance damping, and we will present the improvements implemented to tackle the previous limitations of the PQCG.

\subsection{Noise and stability issues}

\subsubsection{Analysis of the noise of the system.}
The SQUID is an ultra sensitive magnetic flux-to-voltage converter, and hence interferring electromagnetic signals are critical: both, low-level high-frequency noise and sharp transients, can be detrimental to the operation of the system. In the PQCG, the noise to which the SQUID is exposed can have different origins: i) direct flux noise picked up by the SQUID, ii) flux noise coming from the currents that circulate in the CCC windings due to Johnson-Nyquist noise emitted by the resistive elements connected to them, and iii) external electromagnetic noise picked-up in the circuit through capacitive or inductive couplings. Moreover, the CCC is a complex system of 15 magnetically coupled superconducting windings with distributed capacitances and with inductances $L_i \sim i^2 L_\mathrm{CCC}$, where $L_\mathrm{CCC}$ is the inductance of the toroidal shield and $i$ the number of turns. For example, for the windings of 2065 turns, $L_\mathrm{2065}$ = 85 mH and the equivalent parallel capacitance is 1 nF. The total flux noise $\sqrt{S^\mathrm{tot}_{\phi}(f)}= \sqrt{S_\mathrm{SQUID}(f)+S_\mathrm{CCC}(f)}$ can be deduced from the power spectral density $S^\mathrm{tot}_{\phi}(f)$, where $S_\mathrm{SQUID}(f)$ and $S_\mathrm{CCC}(f)$ are the respective contributions from the noise of the SQUID (intrinsic or captured), and from the noise of the CCC and the circuit connected to it. The latter is the most difficult to analyse. Thus, to have a better understanding, a model of the circuit is useful. Here we have used a simplified model of two magnetically coupled series-RLC circuits as a generic model to account for the main features of the spectra (see section \ref{model}) in different situations. Because it includes the magnetic coupling between the windings, this model goes beyond our previous simulations \cite{Azib2018}.

\begin{figure*}[!h]
  \centering
  \includegraphics[width=4in]{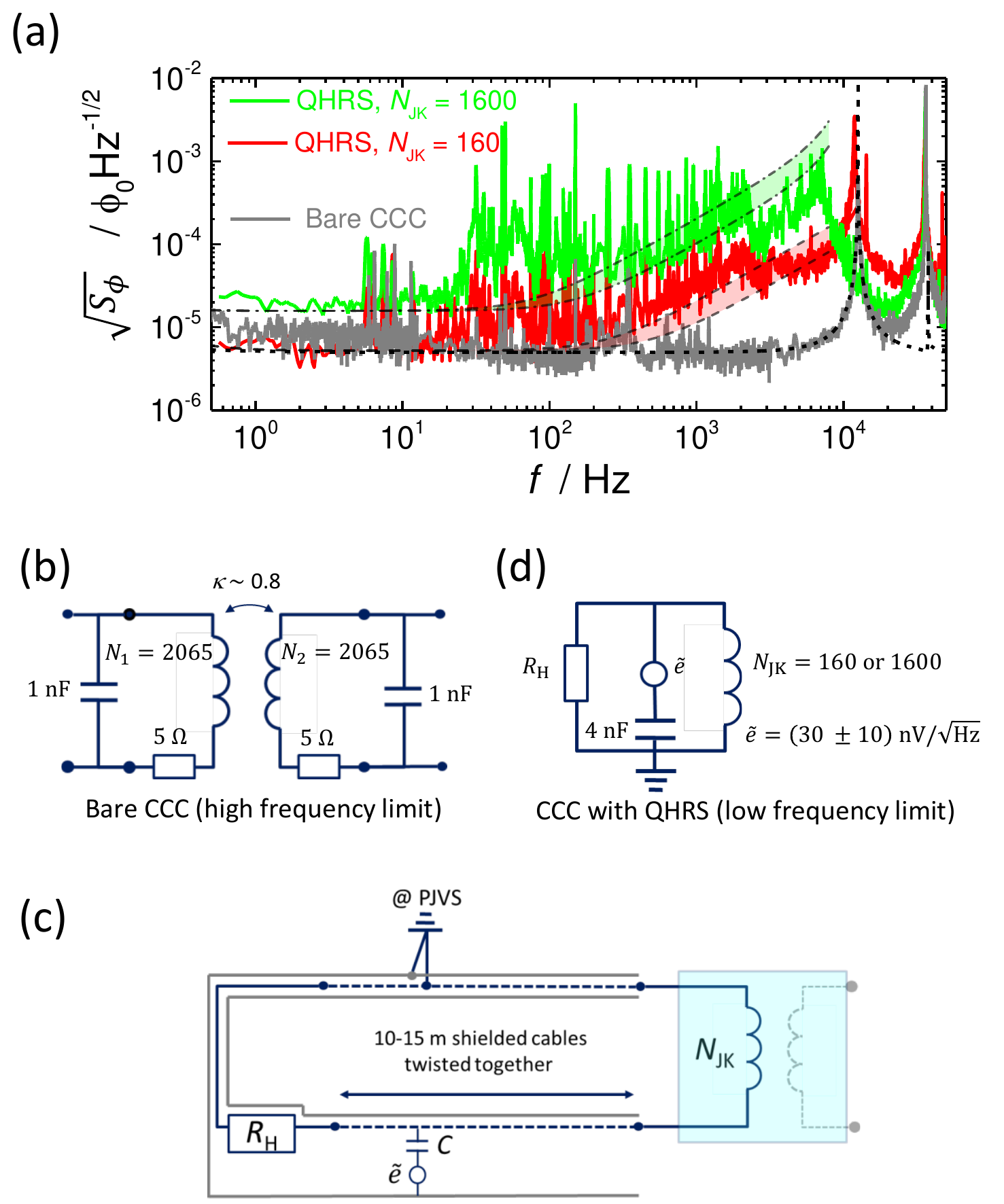}
  \caption{\small{(a) Comparison of the flux noise measured at the output of the SQUID in internal feedback mode: for the bare CCC (grey line) and for the CCC doubly connected to the QHRS through the windings of 160 (red line) and 1600 (green line) respectively. Black lines are simulations calculated from the expression $S^\mathrm{tot}_{\phi}(f)= S_\mathrm{SQUID}(f)+S_\mathrm{CCC}(f)$, with $S_\mathrm{SQUID}(f) = A^2 (f_c/f)+B^2$ ($A=$ 3 $\mu\phi_{0}/\sqrt{\mathrm{Hz}}$, $B=$ 5 $\mu\phi_{0}/\sqrt{\mathrm{Hz}}$ , $f_\mathrm{c}$ = 0.3 Hz). $S_\mathrm{CCC}(f)$, in the case of the bare CCC (black short-dashed line),  is calculated from the model presented in (b) of two magnetically coupled RLC-circuits and the corresponding parameters (see section \ref{model}). In the case of the CCC connected to the QHRS, for $N_\mathrm{JK}$ = 160 (black dashed line) and 1600 (black dash dotted line), $S_\mathrm{CCC}(f)$ is calculated from the model presented in (c) and (d). (c) A schematic representation of the connection of one winding to the QHRS through the long shielded cables. The capacitance to ground and the voltage noise source are represented. (d) The model for the CCC connected to the QHRS used to account for the low frequency part of the spectrum is a RL-circuit in parallel with a capacitance to ground of 4 nF and a noise source of amplitude $\tilde{e}$=(30 $\pm$ 10) nV$/\sqrt{\mathrm{Hz}}$.}}
  \label{fig:noise}
\end{figure*}

Figure\space\ref{fig:noise}(a) compares the noise spectra of the CCC not connected (bare CCC) and connected to the QHRS, without the damping circuit. For the present study, we have performed noise measurements on the output voltage of the dc-SQUID when operating it in internal feedback mode and with the highest bandwidth (0.75 V/$\phi_0$ sensitivity and 50 kHz bandwidth). We present the results in terms of the flux noise, $\sqrt{S^\mathrm{tot}_{\phi}(f)}$, expressed in $\phi_0/\sqrt{\mathrm{Hz}}$.

For the bare CCC, two resonances appear at 12.6 kHz and 36.4 kHz. They can be identified as the LC-resonances resulting from the magnetic coupling of the two identical windings of 2065 turns short-circuited by their parasitic capacitances (see figure\space\ref{fig:noise}(b) and annex \ref{model}). Considering that the mutual inductance is given by $M_{2065} = \kappa L_{2065}$, where $\kappa<1$ is the coupling factor, the assignment of the LC-resonances to the calculated frequencies $f_0/\sqrt{1-\kappa}$ and $f_0/\sqrt{1+\kappa}$, where $f_0$ is the LC-resonance without coupling, leads to $f_0$ = 17.2 kHz and $\kappa\sim 0.8$, which are reasonable values for the CCC \cite{Rengnez_2015}. The calculated noise spectrum of figure\space\ref{fig:noise}(a) is in good agreement with the experimental noise spectrum of the bare CCC down to 10 Hz.  At lower frequencies, the noise level can be explained by the dominant contribution of the SQUID noise with a slightly increased white noise level of 5 $\mu\phi_{0}/\sqrt{\mathrm{Hz}}$ compared to the SQUID's intrinsic noise.

The noise spectra when the QHRS is doubly connected to the windings of 160 and 1600 (red and green lines of figure\space\ref{fig:noise}(a)) have been measured to clarify the problem of the reduced stability when increasing $N_\mathrm{JK}$. In this case, at low frequency, the bottom white noise levels are well explained by considering both contributions of the SQUID noise and the Johnson-Nyquist noise of the QHRS $\frac{N_\mathrm{JK}}{\gamma_{CCC}}\sqrt{\frac{4 k T}{R_\mathrm{H}}}$, where $k$ is the Boltzmann constant and $T$ the temperature. The latter becomes dominant for 1600 turns and amounts to 15 $\mu\phi_0$/$\sqrt{\mathrm{Hz}}$. To avoid this increased noise, an optimum value of $N_\mathrm{JK}$ would be about 500 turns. However, the most striking feature is the almost linear increase from 100 Hz up to 10 kHz, on which 50 Hz harmonics are superimposed. This extra noise, the magnitude of which scales with $N_\mathrm{JK}$, is explained by a coupling of the circuit with external electrical noise sources. It can be modeled by a voltage noise source coupled to the circuit through the capacitance to ground of the 15-m long shielded twisted pairs connecting the CCC and the QHRS \cite{Remark1} as shown schematically in figure\space\ref{fig:noise}(c). This capacitance has been measured and amounts to 4 nF. To account for the increased noise in the frequency range 100 Hz to 10 kHz, we have simulated the quantum circuit as a simple RL-circuit considering a single winding connected to the QHRS. A 4-nF capacitance in series with a noise source is connected in parallel to the inductance of the winding as shown in figure\space\ref{fig:noise}(d). The red and green hatched areas in figure\space\ref{fig:noise}(a) correspond to adjustments of the effective white noise voltage source to (30 $\pm$ 10) nV$/\sqrt{\mathrm{Hz}}$, for both $N_\mathrm{JK}$ = 160 and 1600. The good agreement confirms that the SQUID instability observed for large values of $N_\mathrm{JK}$ comes from the existence of external noise sources that couple to the quantum circuit and excite the high-frequency resonances especially as large $N_\mathrm{JK}$ values are used. Moreover, when the battery powered programmable Josephson bias source is connected, its additional digital noise is sufficient to cause the system to become unstable even in internal feedback mode. In future we will improve this situation by reducing the length of the cable between the CCC and the QHRS. For the moment this length is fixed by the current location of the cryostats in the laboratory.



\subsubsection{Damping.}

\begin{figure*}[!h]
  \centering
  \includegraphics[width=3.5in]{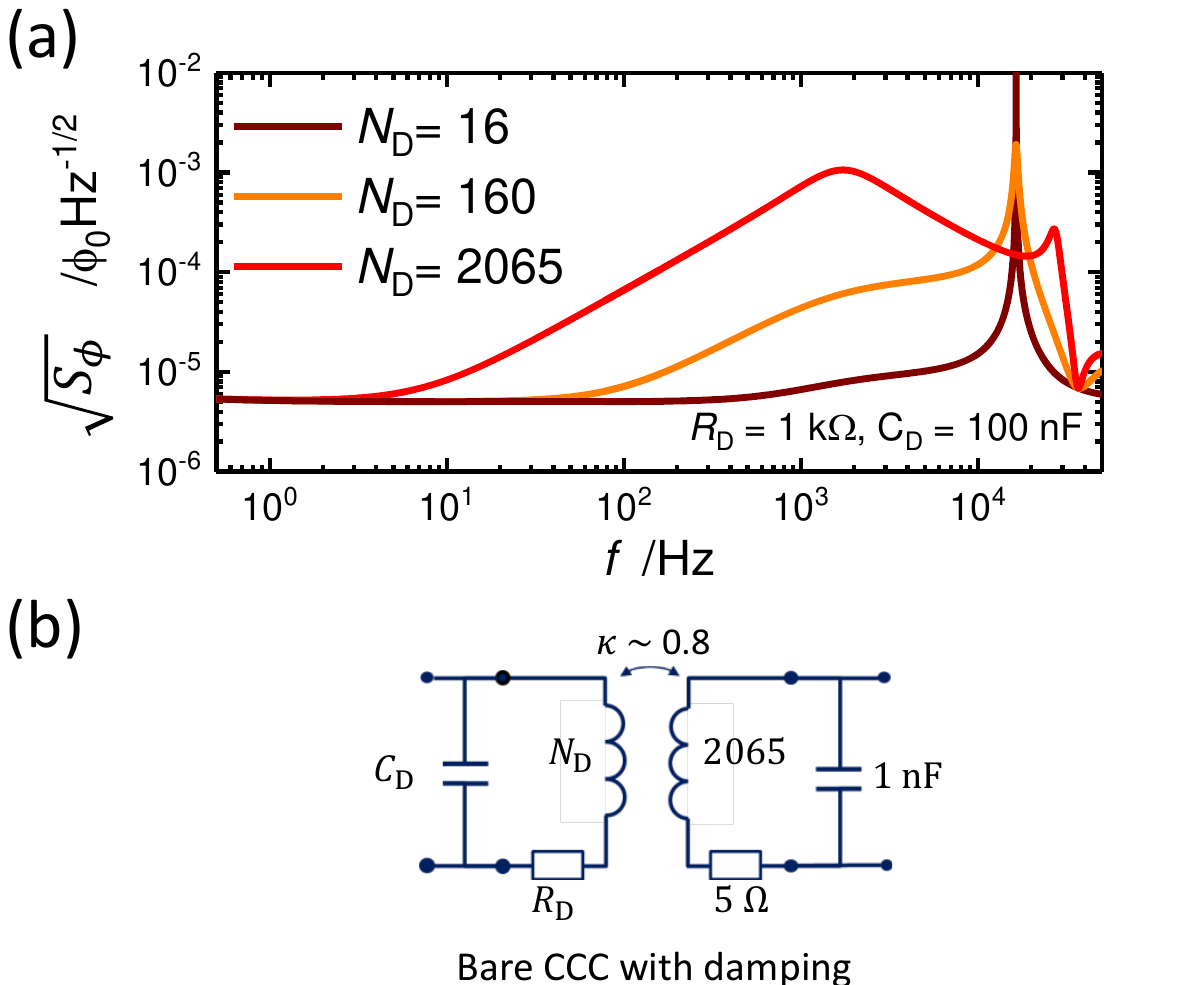}
  \caption{\small{(a) Simulated flux noise for the bare CCC in the presence of the damping circuit, for different $N_\mathrm{D}$= 16, 160 and 2065, with $R_\mathrm{D}=1$ k$\Omega$ at $T_\mathrm{D}=$ 293 K and $C_\mathrm{D}$ = 100 nF. The simulation is based on the model presented in (b). (b) Magnetic coupling of a winding of 2065 turns with the damping circuit (section \ref{model}).}}
  \label{fig:damping}
\end{figure*}

Given the coupling of external noise to the PQCS, it was therefore crucial to damp the high frequency resonances \cite{Brun-Picard2016} to stabilize the SQUID operation. The main question is how to choose the parameters of the damping. Experimentally, we observed that high values of $N_\mathrm{D}$ enhanced the effect of the damping. The results of a simulation shown in figure\space\ref{fig:damping}(a) illustrates the increasing efficiency of the damping on the high frequency peak for three different values of $N_\mathrm{D}$ from 16 to 2065. The simulation is based on the same generic model for the CCC, i.e. the series coupled RLC-circuit of in figure\space\ref{fig:noise}(b), in which the values of the resistance and the capacitance of one of the circuits have been replaced by the values used in reference \cite{Brun-Picard2016}, $R_\mathrm{D}$ = 1 k$\Omega$ and $C_\mathrm{D}$= 100 nF (see figure\space\ref{fig:damping}(b)).

\begin{figure*}[!h]
  \centering
  \includegraphics[width=3.5in]{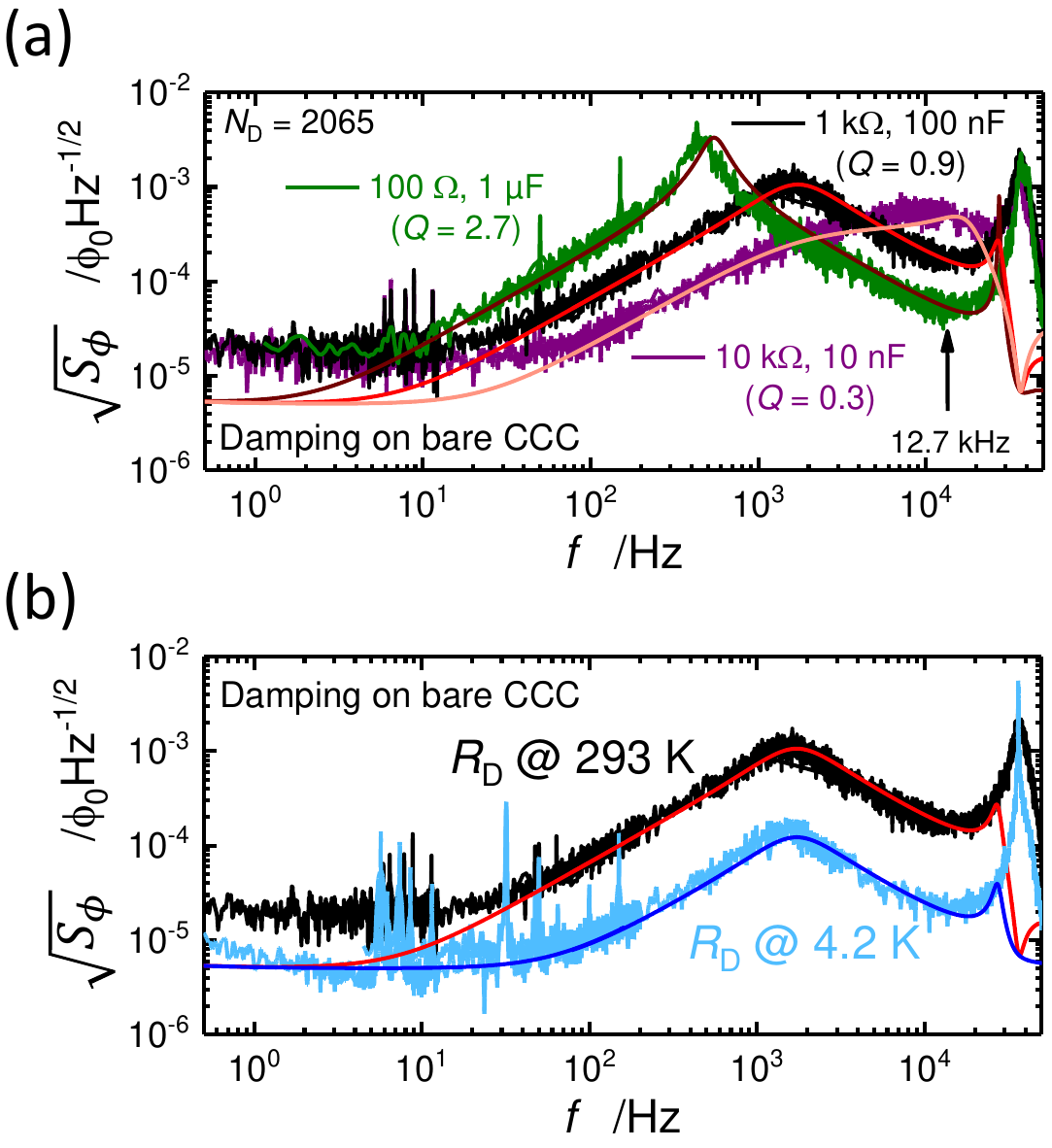}
  \caption{\small{Experimental data for the damping : (a) Flux noise for different sets of damping parameters [$R_\mathrm{D}$,$C_\mathrm{D}$]. The resistor $R_\mathrm{D}$ is placed at ambient temperature. The three configurations, [100 $\Omega$, 1 $\mu$F], [1 k$\Omega$, 100 nF] and [10 k$\Omega$, 10 nF] respectively (dark green, black and purple lines, respectively), correspond to the same RC-product but to different quality factors $Q$ from 2.7, 0.9 and 0.3, respectively. The red-gradient lines (from dark to light) represent the calculated $\sqrt{S_{\phi}^\mathrm{tot}(f)}$ for $S_\mathrm{CCC}(f)$ based on the model in figure\space\ref{fig:damping}(b) for the corresponding [$R_\mathrm{D}$,$C_\mathrm{D}$] parameters. (b) Comparison of the flux noise for $R_\mathrm{D} =$\space1\space k$\Omega$ at ambient temperature (black line) and at 4.2 K (light blue line). The red and blue lines are the corresponding calculated spectra.}}
  \label{fig:damping2}
\end{figure*}

Since the first implementation of the damping circuit in the PQCG with $N_\mathrm{D}$ = 1600 \cite{Brun-Picard2016}, the winding of 2065 turns has then been favored \cite{Azib2018} not only to achieve a better efficiency but also to allow the use of the 1600-turns winding for $N_\mathrm{JK}$. At intermediate frequency (100 Hz $<f<$ 10 kHz), the damping circuit is responsible for a damped resonance much broader than the self-resonance of the bare CCC \cite{Drung2009,Brun-Picard2016,Williams2020}. As a first approximation, the center of the broad peak is at $f_\mathrm{D}\sim1/(2\pi \sqrt{L_\mathrm{D} C_\mathrm{D}}$) and its amplitude, due to the Johnson-Nyquist noise of $R_\mathrm{D}$, is given by $\sim\frac{N_\mathrm{D}}{\gamma_\mathrm{CCC}}\sqrt{\frac{4 k T_\mathrm{D}}{R_\mathrm{D}}}$ \cite{Brun-Picard2016}. Once the winding has been chosen, $N_\mathrm{D}$ is fixed and hence $L_\mathrm{D}$ is fixed. The choice of the capacitance $C_\mathrm{D}$ is made such that the broad peak is well below 10 kHz so that the damping is efficient at 12 kHz. The last parameter, $R_\mathrm{D}$, fixes the amplitude of the broad resonance and the quality factor $Q \sim \frac{1}{2 \pi f_\mathrm{D} R_\mathrm{D} C_\mathrm{D}}$. Noise measurements have been performed for the bare CCC for three different sets of damping parameters [$R_\mathrm{D}$,$C_\mathrm{D}$] ($R_\mathrm{D} C_\mathrm{D} = 10^{-4}$ s), with $N_\mathrm{D}$ = 2065 (see figure\space\ref{fig:damping2}(a)). Three situations are represented corresponding to $Q \sim$ 2.7, 0.9 and 0.3 respectively. The flux noise has been calculated, for the corresponding parameters, in the case of the damping circuit magnetically coupled to a winding of 2065 turns (figure\space\ref{fig:damping2}(b), and they account very well for the complex shapes of the spectra in the three cases.

Finally, we have chosen $R_\mathrm{D} = 1$ k$\Omega$ and $Q\sim1$ as a good compromise because it suppresses the high quality factor resonance at 12.6 kHz, such that the main goal has been reached, and it limits the amplitude of noise towards the low frequency part. However, the amplitude of the broad peak is about 1 m$\phi_0/\sqrt{\mathrm{Hz}}$. Two possible improvements can be considered, increasing of the resistance by a factor of 2 to reach the critical value $Q=1/2$ and/or placing the resistance at 4.2 K. The latter point will be detailed in the present paper. Note that the resonance at 35 kHz did not disappear, it has been broadened and slightly shifted to 36 kHz; for the moment, this is not clearly understood, and the model cannot account for it because only two windings are simulated. Its possible detrimental effect on the stability may be limited anyhow by the damping, which attenuates part of the high-frequency components of the signal during the fast reversals that are likely to excite this resonance.


More unexpectedly, figure\space\ref{fig:damping2}(a) shows that the low frequency part of the spectrum is enhanced from about 7 to 20 $\mu\phi_{0}/\sqrt{\mathrm{Hz}}$ at 1 Hz compared to the bare CCC. This may be a consequence of mixing-down effects due to the high level of broadband noise in the kHz range. The SQUID is used in flux-locked loop (FLL) in order to amplify and linearize the periodic $V-\phi$ transfer function of the SQUID. The SQUID is then locked at a given working point of the $V-\phi$ characteristic by the feedback loop. At low frequencies, within the FLL bandwidth, the FLL suppresses the effect of the noise on the SQUID, in other words, the flux 'seen by the SQUID' is constantly compensated by FLL. However, if the noise above the FFL bandwidth is sufficiently strong, then the linearized flux range can be exceeded, leading to non linear distortions, and consequently to rectified noise which will be superimposed with the low frequency feedback signal. Hence these mixing down effects deteriorate the signal-to-noise ratio \cite{Drung1996} but, unless the amplitude of the noise is much higher, the SQUID is still locked at its working point.

This hypothesis has been confirmed recently, as the noise level at low frequency retrieves the expected one, as soon as the peak in the intermediate bandwidth is reduced by a factor of 8.5 by placing the resistance $R_\mathrm{D}$ = 1 k$\Omega$ at 4.2 K, as shown in figure\space\ref{fig:damping2}(b). The integrated noise between 100 Hz and 10 kHz can be used to quantify the detrimental noise level. For instance, it is equal to 60 m$\phi_{0}$ for $R_\mathrm{D}$ at ambient temperature (black line), whereas it falls at 7 m$\phi_{0}$ for $R_\mathrm{D}$ at 4.2 K (light blue line).

\subsubsection{External feedback.}

In external feedback mode, a CCC winding is used as the feedback path instead of the dedicated modulation coil of the SQUID (internal feedback mode). It has a big impact on the dynamic of the system and the problems of stability are even more severe.

Compared to the internal feedback, the stability of the FLL is ensured by choosing an operation mode (5S) with a reduced bandwidth of 500 Hz as suggested in the documentation of the SQUID \cite{SQUID_QuantumDesign_550}, i.e. characterized by a cutoff frequency much lower than the high frequency resonances. However, despite both this change and the fine adjustment within $10^{-6}$ of the current ratio $I_\mathrm{PQCG}/I_\mathrm{PQCS}$, the system proved to be unstable (i.e. frequent unlocking of the SQUID) for high ampere-turns reversals, which are needed during the PQCG operation to eliminate voltage offsets. This instability is related to the periodic nature of the $V-\phi$ characteristic of the SQUID, with several equivalent working points, each separated by a flux quantum. Fast reversals of the signal can force the SQUID to jump between adjacent working points if large signals are applied and if they change faster than the FLL can track. These problems are very similar to those encountered in different CCC resistance bridges \cite{Sanchez2009,Drung2009,Poirier2020}, where two current sources feed the resistors to be compared, each being connected in series with a ratio winding of the CCC. A special care is paid to keep the bridge balanced during current reversals, by matching the filters on both sides of the bridge \cite{Sanchez2009, Williams2011} and by using offset, in-phase and in-quadrature correction circuits \cite{Poirier2020}. Other authors have implemented wideband feedback \cite{Drung2009}. However, in the case considered here, because the two circuits connected to the CCC, the PQCS circuit and the output circuit, are biased by a voltage source and a current source, respectively, these adjustments are more difficult.

\begin{figure*}[!h]
  \centering
  \includegraphics[width=4.5in]{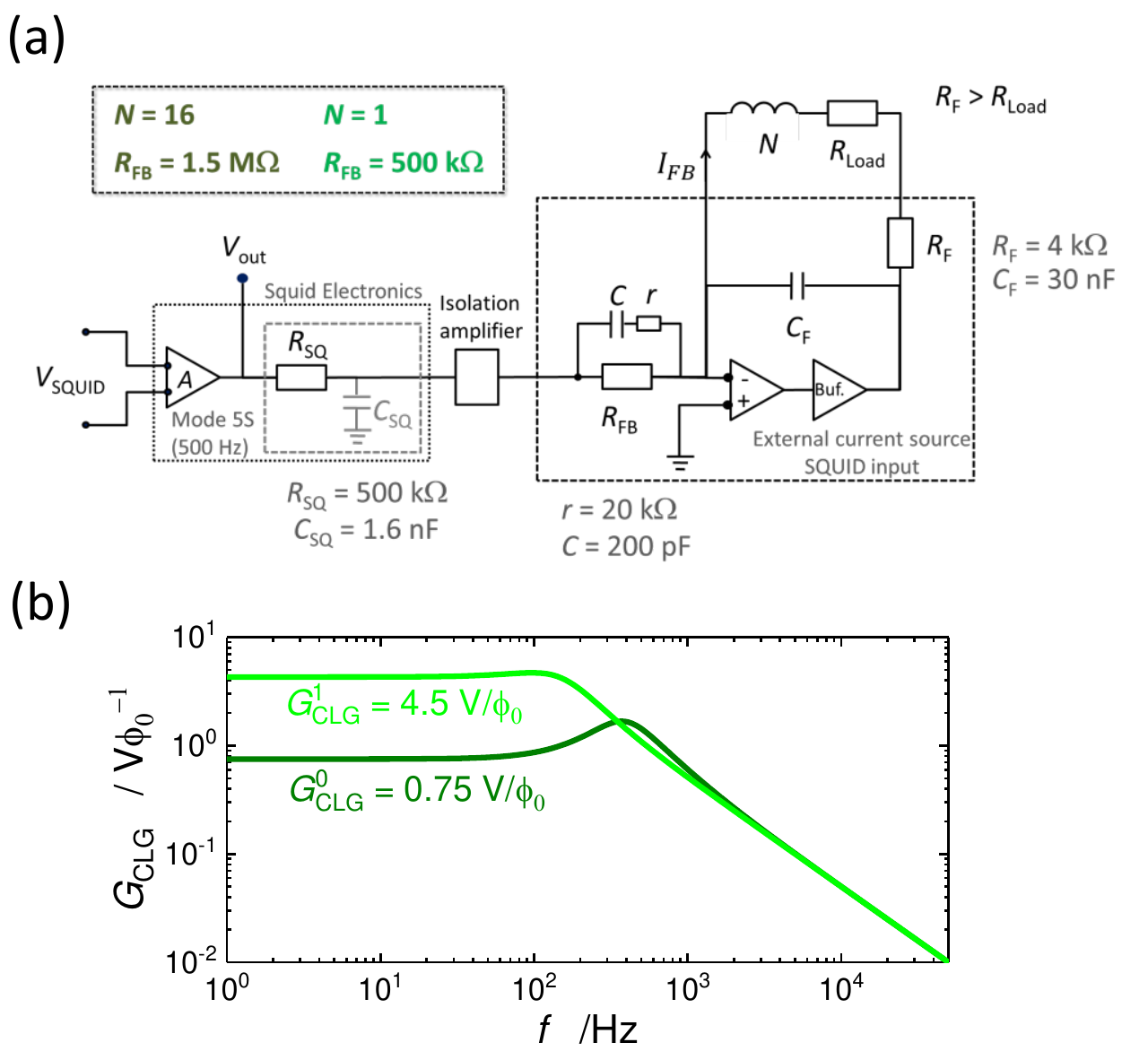}
  \caption{\small{(a) Electronic circuit of the external feedback : details of the SQUID input stage of the external current source. The different parameters used for the calculation of the gain of the external feedback are given. (b) Frequency response of the external feedback loop calculated for the two different closed-loop gain $G_\mathrm{CLG}^0=$ 0.75 V/$\phi_{0}$ and $G_\mathrm{CLG}^{1}=$4.5 V/$\phi_{0}$, corresponding to $R_\mathrm{FB}$ = 1.5 M$\Omega$ for $N$=16 and $R_\mathrm{FB}$ = 500 k$\Omega$ for $N$=1.}}
  \label{fig:EFB}
\end{figure*}

This instability has been overcome by modifying the closed-loop feedback gain. Let us detail the original external feedback circuit as illustrated in figure\space\ref{fig:EFB}(a). The SQUID output voltage (mode 5S, bandwidth 500 Hz) is connected to the input of a high impedance differential amplifier to avoid loading the 500 k$\Omega$ resistance at the output of the SQUID electronics with the capacitance of the long cable plugged into the external current source. However a parasitic capacitance $C_\mathrm{SQ}$ of 1.6 nF loading the resistance has been measured at the output of the circuit connecting the SQUID to the isolation amplifier. The same closed-loop feedback gain of $G_\mathrm{CLG}^0$ = $R_\mathrm{FB} \frac{\gamma_\mathrm{CCC}}{N}$= 0.75 V/$\phi_{0}$ as in the internal feedback mode of the SQUID has been targeted for a winding of $N$=16 turns; therefore, a feedback resistance $R_\mathrm{FB}$ = 1.5 M$\Omega$ has been chosen. A small capacitance of 200 pF in series with a 20 k$\Omega$ resistor is connected in parallel to $R_\mathrm{FB}$ to partially compensate the phase shift caused by the 1-kHz low-pass filter ($R_\mathrm{F}$ and $C_\mathrm{F}$) present at the output of the current source \cite{Poirier2020}. Nevertheless, a residual phase shift exists in the feedback circuit \cite{Drung2004} that translates into a peak in the frequency response of the feedback loop (dark green line of figure\space\ref{fig:EFB}(b)), according to the calculated response based on the circuit presented in figure\space\ref{fig:EFB}(a). From the simulation, it can be deduced that the phase shift comes mainly from the parasitic capacitance, $C_\mathrm{SQ}$. This certainly explains the instability against current reversals in the PQCG. 
By increasing the closed-loop gain to 4.5 V/$\Phi_0$ for the PQCG, the peak in the frequency response has almost disappeared, as shown by the green line in figure\space\ref{fig:EFB}(b). This smooth response of the feedback with a slightly reduced bandwidth has solved the problem of stability against high ampere-turns reversals. In the future, a possible improvement would be to reduce the parasitic capacitance. However, the impact of the higher closed-loop gain on the accuracy of the generated current has been tested in \cite{Brun-Picard2016} and it revealed negligible.

\subsection{Set-up improvements.}\label{Improvements}
To reduce the common mode noise on the PQCS circuit, a common mode torus has been placed on the output of the PJVS, on the current leads that supply the QHRS, as depicted in figure\space\ref{fig:schema}. A similar solution had been implemented on the output of the voltage-controlled current source \cite{Poirier2020}. Guided by the reduction of the noise at frequencies below 10 Hz while ensuring the stability even for the highest ampere-turns, i.e. for $N_\mathrm{JK}$ = 160 and 1600 when the CCC is connected to the PQCS, we found that the noise and therefore the stability of the PQCG is very sensitive to the position of the ground connection. In \cite{Brun-Picard2016}, in order to reduce the effect of leakage currents redirected to ground, it was connected at 4.2 K at the low voltage pad of the Josephson array and on the current contact of the QHRS. However, the relative error due to the leakage currents has been estimated to be less than $10^{-11}$, and moving the position to room temperature does not change much from this point of view. On the contrary, the ground connection at room temperature allows testing different positions and their effect on the noise of the circuit. We found that the best position, reducing the low frequency noise from 25 to 7 $\mu$V/$\sqrt{\mathrm{Hz}}$, is at the head of the CCC as shown in figure\space\ref{fig:schema}, on the voltage contact of the QHRS. This certainly short-circuits a noise source ($\tilde{e})$ that couples directly into the windings. The presence of the high inductance of the common mode torus on the current lead might explain that the noise couples less on this lead.

\begin{figure*}[!h]
  \centering
  \includegraphics[width=3.2in]{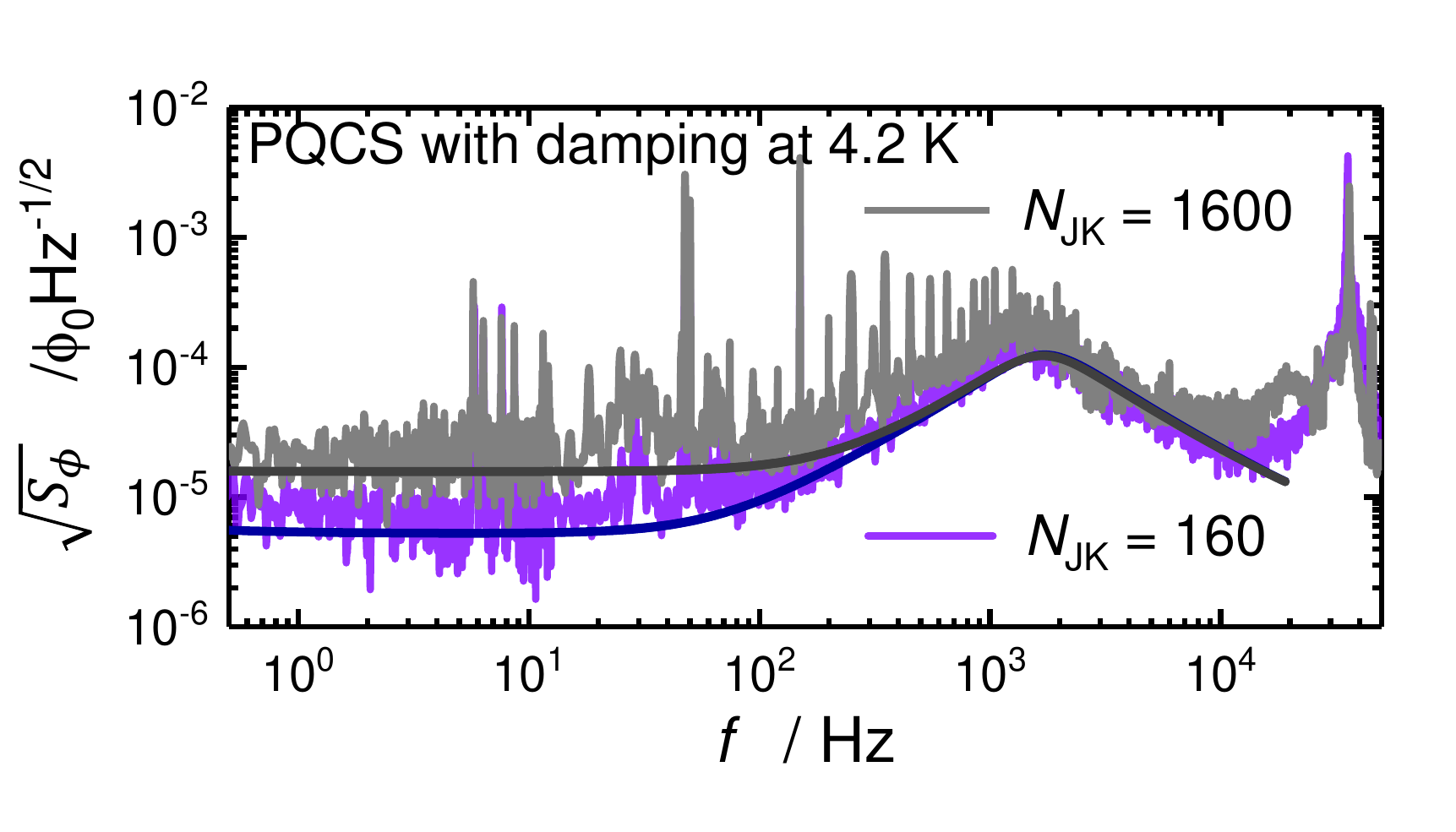}
  \caption{\small{Flux noise of the PQCS connected on $N_\mathrm{JK}$ = 160 and 1600, with the damping circuit $R_\mathrm{D}$ = 1 k$\Omega$ at 4.2 K and $C_\mathrm{D}$ = 100 nF. The simulations are calculated for the circuit of figure\space\ref{fig:damping}(a), where the parameters of the right loop are $R_\mathrm{H}$, $C \rightarrow \infty$, and $L_{160}$ (dark blue line) or $L_{1600}$ (dark grey line).}}
  \label{fig:Fig_modif_PQCS}
\end{figure*}

However, in this configuration, the amplitude of the voltage steps of the PJVS were reduced due to an increased noise that perturbed the array. This has been resolved by adding a 1 $\mu$F highly isolated capacitance placed between the 0-V of the battery powered Josephson programmable bias source and the ground (see figure\space\ref{fig:schema}). This new configuration is much more reliable, the array is less subject to flux trapping than it was previously with the ground connection at low temperature. Moreover, a heating system has been implemented at PTB on the Josephson array chip (see figure\space\ref{fig:schema}), and allows to get rid of trapped flux within few minutes.

Finally, after all these modifications, the system has gained stability in internal feedback mode. As shown in figure\space\ref{fig:Fig_modif_PQCS}, the PQCS (including the PJVS bias source) can be connected notably to $N_\mathrm{JK}$ = 160 (purple line) and 1600 (grey line) and the low frequency noise in both configurations is close to the expected levels. The flux noise of the broad resonance integrated between 100 Hz and 10 kHz is below 10 m$\phi_0$ in both cases which is close to the 7 m$\phi_0$ for the bare CCC (see figure\space\ref{fig:damping2}(b)). However, in external feedback mode, the stability issues still limit the operation to $N_\mathrm{JK}$ = 160.

Compared to reference \cite{Brun-Picard2016}, not only the noise in the intermediate frequency range has been reduced but also the low-frequency noise below 10 Hz. The reduction of the latter by a factor of 2 and the increase of $N_\mathrm{JK}$ by a factor of 1.25, may lead to a maximum increase of the signal-to-noise ratio by a factor of 2.5 for the most accurate measurements. However, this potential gain may be reduced if the excess noise ($1/f$ noise) becomes the dominant noise at very low frequencies \cite{1_over_f_noise}. Note that the PQCG experiments aiming for highest accuracy are performed at current-reversal frequencies below 100 mHz. Hence, new measurements are needed to confirm the improvement in Type A uncertainty of the PQCG \cite{Brun-Picard2016}. Nonetheless, the reduction of the noise in the intermediate frequency range has already improved the Type A uncertainty of digital ammeter calibrations as will be seen in the next section.

\section{Digital ammeter calibrations}\label{DA_calibration}

Here we present the calibrations of two precision digital ammeters (DA), an HP3458A (DA$\sharp$1) and a Fluke 8508A (DA$\sharp$2), in the mA ranges. During the calibrations, the ammeter replaces $R_\mathrm{Load}$ in figure\space\ref{fig:schema}. However to avoid perturbing the SQUID with the digital noise from the DA, a highly insulated PTFE capacitor (100 nF) is added at the current measurement terminals. The low-potential input is connected to ground. Prior to the calibration campaign, DA$\sharp$1 has been adjusted by using a 10-k$\Omega$ resistor standard and a 10-V Zener voltage standard calibrated in terms of $R_\mathrm{K}$ and $K_\mathrm{J}$, respectively. DA$\sharp$2 has not been adjusted.

\begin{figure*}[!h]
  \centering
  \includegraphics[width=4.2in]{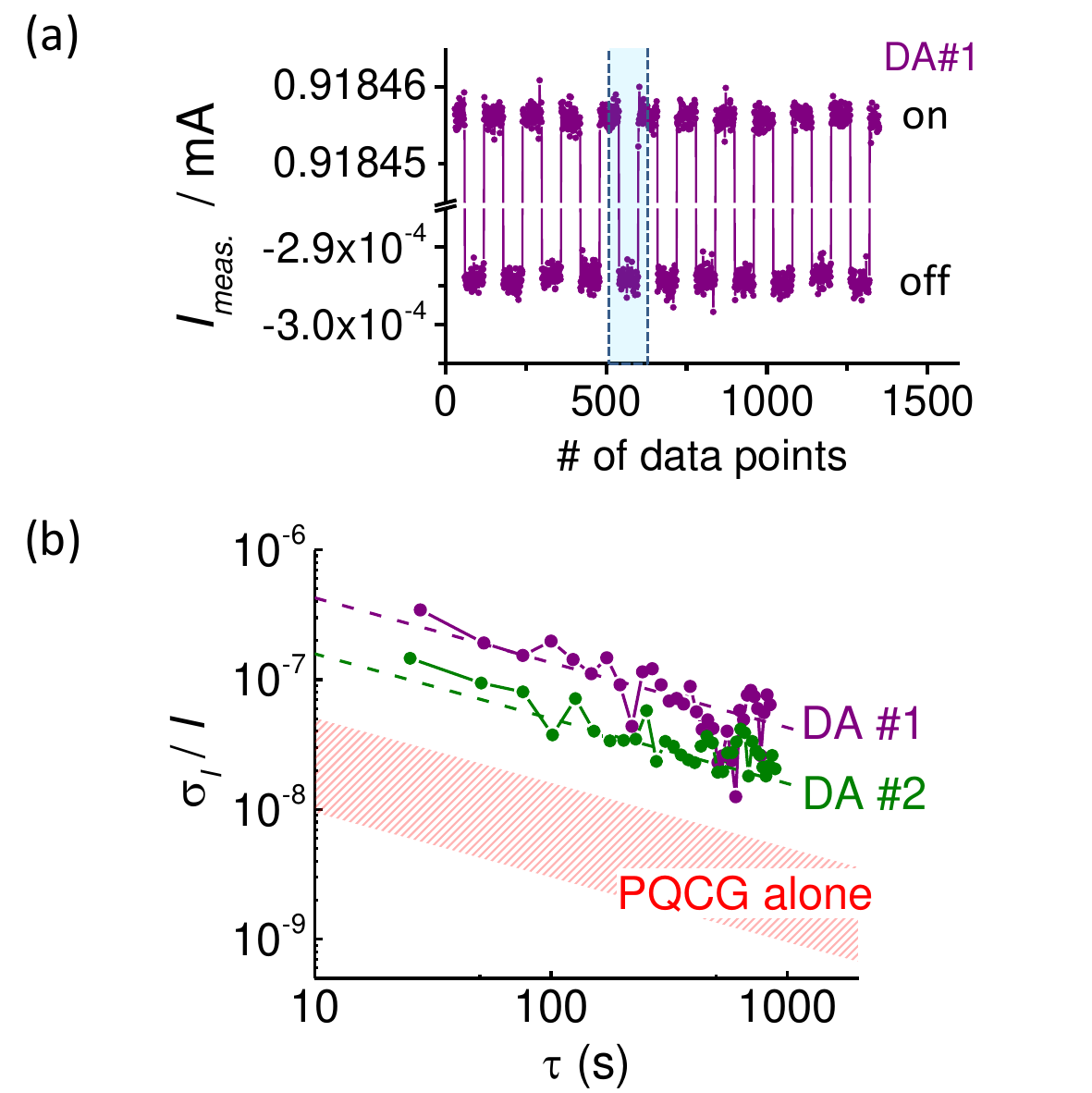}
  \caption{\small{(a) Current reversals for DA$\sharp$1 at 0.918 mA, the blue area highlights the data of one on-off-on cycle. (b) Relative Allan deviations, $\sigma_I/I_\mathrm{[DA\sharp1]}$ and $\sigma_I/I_\mathrm{[DA\sharp2]}$, calculated from 71 on-off-on cycles at\space0.918 mA ($N$ = 4) for DA$\sharp$1 (purple dots) and DA$\sharp$2 (olive dots), respectively. $\tau_0$ = 28 s and 25.4 s for DA$\sharp$1 and DA$\sharp$2, respectively. The $\tau^{-1/2}$ fits (purple and olive dashed lines respectively) reveal a white noise regime over about 800 s. The red hashed area represents the relative Allan deviation $\sigma_I/I_\mathrm{[PQCG]}$ expected for the PQCG for the same current reversal procedure. It is  calculated from the expression of the Allan deviation $\sigma_I/I_\mathrm{[PQCG]}(\tau) = A'_\mathrm{rel}/\sqrt{n_\mathrm{s} \tau_0}$ (see section \ref{Calc_Avar}), where $n_\mathrm{s}$ is the number of cycles and where $A'_\mathrm{rel}$ is a constant fixed by the relative standard deviation for one cycle $\sigma_\phi/\phi_\mathrm{[PQCG]}(\tau_0) = \sqrt{2S_\phi/\tau_\mathrm{e}}/\phi$, where $\tau_\mathrm{e}$ is the effective measuring time over the cycle, disregarding the waiting time, and for $\sqrt{S_\phi}$ spanning from 8 to 50 $\mu\phi_0/\sqrt{\mathrm{Hz}}$.}}
  \label{fig:Fig_Cal_3458-2}
\end{figure*}

Before each calibration, we systematically verify the quantization of the PJVS by checking the current amplitudes of the quantized voltage steps, $U_\mathrm{J} = 0, \pm n_\mathrm{J} f_\mathrm{J} K^{-1}_\mathrm{J}$, of each segments at the operating frequency and at the optimum microwave power. The relative correction $\alpha = 1.6\times10^{-7}\pm2.5\times10^{-9}$ for $N_\mathrm{JK}$ =160 is calculated by taking into account the resistances of the double connection.

Each calibration point takes less than 10 minutes. An on-off-on cycle is used to subtract the offsets, the measured value is calculated from eleven cycles as shown in the figure\space\ref{fig:Fig_Cal_3458-2}(a). We tried to realize as much as possible similar acquisition times for the two DAs despite the different parameter settings and the different dead times (see section \ref{Calc_Avar}). The duration $\tau_0$ of an on-off-on cycle is 52 s for DA$\sharp$1 (and 35.6 s for DA$\sharp$2). A delay $\tau_w$ of 2 s (5 s) after each reversal has been used in order to let the system stabilize.

The relative Type A standard uncertainty of each calibration point is evaluated by experimental standard deviations of the mean. This is justified as the on-off-on cycle is effective in getting rid of the offsets and in making the data sets uncorrelated. This can be seen in figure\space\ref{fig:Fig_Cal_3458-2}(b) from the relative Allan deviations \cite{Witt2005}, $\sigma_I/I_\mathrm{[DA\sharp1]}$ and $\sigma_I/I_\mathrm{[DA\sharp2]}$, for two time series realized with $\tau_0$ = 28 s and 25.4 s for DA$\sharp$1 and DA$\sharp$2, respectively (see section \ref{Calc_Avar} for comments). Both $\sigma_I/I_\mathrm{[DA]}$ show a $\tau^{-1/2}$ dependence on the sampling time $\tau = n_\mathrm{s} \tau_0$ (with $n_\mathrm{s}$ the number of cycles) which reveals a white noise regime over $\sim$ 800 s. The Allan deviations in the experimental conditions of figure \ref{fig:Fig_Cal_3458-2}(b) correspond to typical Type A standard uncertainties of $\sim 6\times10^{-8}$ and $2.3\times10^{-8}$ at $\tau =$ 500 s for DA$\sharp$1 and DA$\sharp$2, respectively. For comparison, the red hashed area corresponds to the relative Allan deviation $\sigma_I/I_\mathrm{[PQCG]}$ expected for the PQCG and calculated for flux noise levels at 20 mHz, spanning from 8 to 50 $\mu\phi_0/\sqrt{\mathrm{Hz}}$ (\cite{1_over_f_noise}, see section \ref{Calc_Avar}) and corresponding to the effective measurement time of the on-off-on sequence of DA$\sharp$1: $\sigma_I/I_\mathrm{[PQCG]}$ is at least about an order of magnitude lower than $\sigma_I/I_\mathrm{[DA\sharp1]}$.

\begin{figure*}[!h]
  \centering
  \includegraphics[width=4.2in]{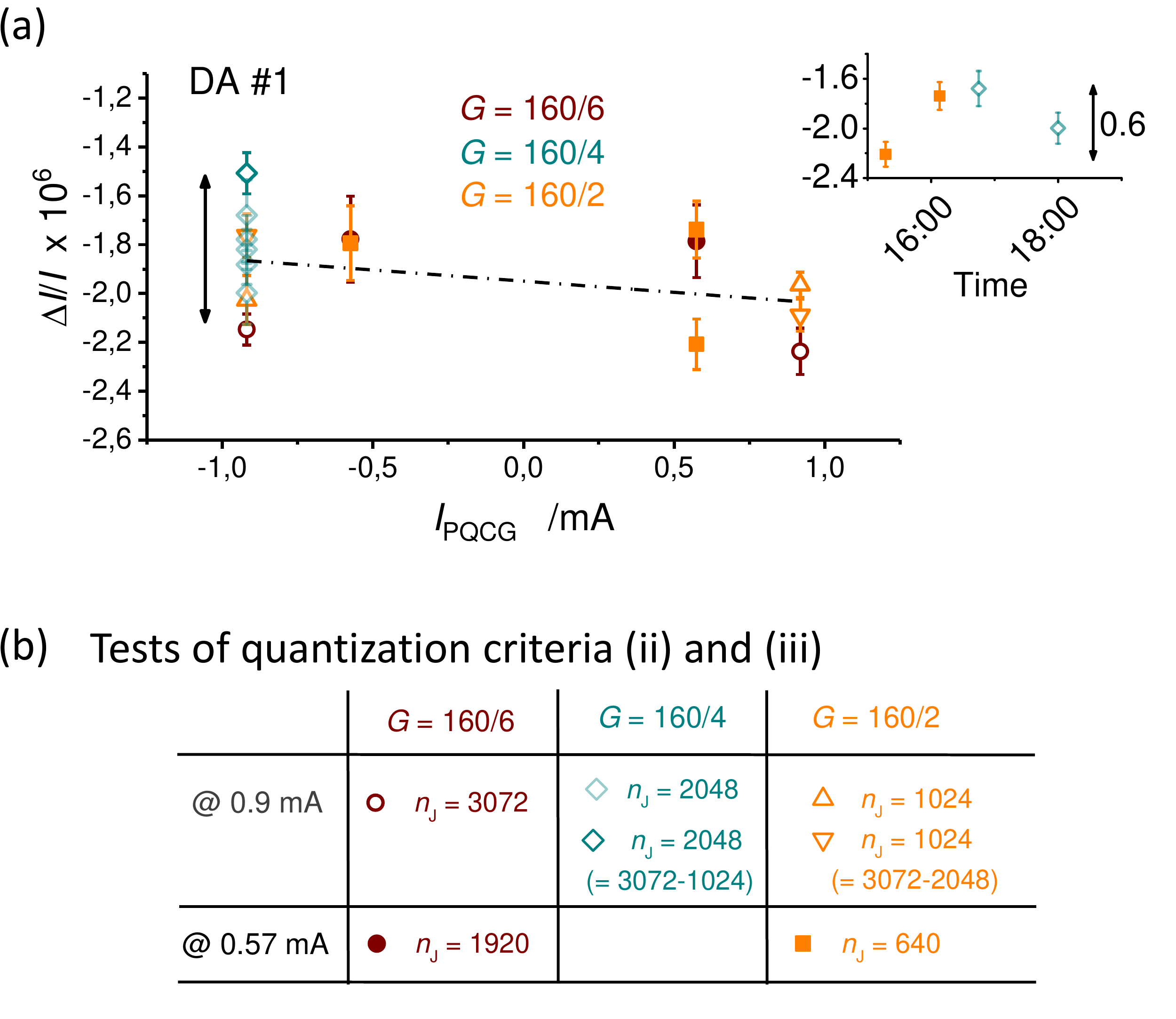}
  \caption{\small{(a) Calibrations of the 1 mA range of DA$\sharp$1 with the PQCG expressed as the relative deviation $\Delta I/I = (I_\mathrm{meas.}-I_\mathrm{PQCG})/I_\mathrm{PQCG}$. The slope of the linear fit (black dot dashed line) is $-0.7\times10^{-7}/$mA. The legend of the different points is detailed in b). Error bars are relative combined uncertainties ($k_\mathrm{c}=1$). The black arrow taken from the inset represents a variation of $\Delta I/I$ of $6\times10^{-7}$. Inset : The repeatability as a function of time for two sets of calibration points (\textcolor[rgb]{1.00,0.50,0.00}{$\blacksquare$}: $n_\mathrm{J}$ = 640 and $G$=160/2 and \textcolor[rgb]{0.00,0.55,0.55}{$\diamond$}: $n_\mathrm{J}$ = 2048 and $G$=160/4) is related to the stability of the DA; a variation of $\Delta I/I$ of the order of $6\times10^{-7}$ can be observed in the same experimental conditions. (b) Quantization criteria for different combinations $[n_\mathrm{J};G]$: Configuration corresponding to the same $n_\mathrm{J}$ but obtained by different segment combinations: \textcolor[rgb]{1.00,0.50,0.25}{$\vartriangle$} and \textcolor[rgb]{1.00,0.50,0.25}{$\triangledown$} correspond to $[1024;160/2]$ and light and dark \textcolor[rgb]{0.00,0.55,0.55}{$\diamond$} correspond to $[2048;160/4]$; Configurations corresponding the same output current: at $\pm$0.9 mA, \textcolor[rgb]{0.50,0.00,0.25}{$\circ$} and \textcolor[rgb]{1.00,0.50,0.00}{\small{$\vartriangle$}} correspond to the configurations [$3072;160/6$] and [$1024;160/2$] respectively; at 0.57 mA, \textcolor[rgb]{0.50,0.00,0.25}{$\bullet$} and \textcolor[rgb]{1.00,0.50,0.00}{$\blacksquare$} correspond to the configurations [$1920;160/6$] and $[640;160/2]$, respectively.}}
  \label{fig:Fig_Cal_3458}
\end{figure*}

\begin{figure*}[!h]
  \centering
  \includegraphics[width=4.2in]{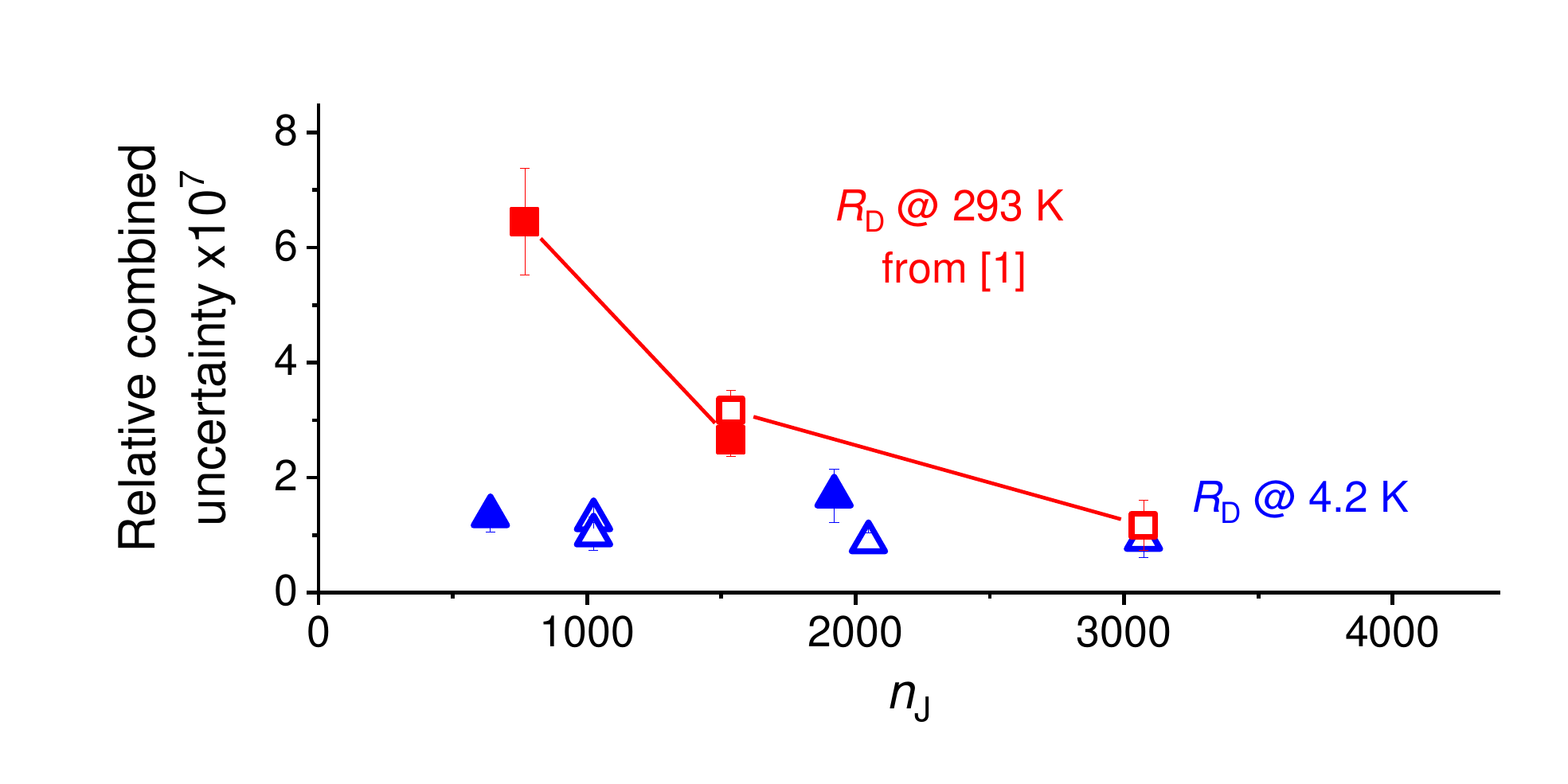}
  \caption{\small{Relative combined uncertainty of the calibration points of DA$\sharp$1 reported in this work (blue triangles) corresponding to the damping resistance $R_\mathrm{D}$ at 4.2 K. Red squares represent the data extracted from Ref.\cite{Brun-Picard2016} corresponding to $R_\mathrm{D}$ at ambient temperature. Open symbols and full symbols correspond to current levels of 0.5 mA and 0.9 mA respectively. Error bars are standard uncertainties (1 s.d.)}}
  \label{fig:Fig_Cal_3458-3}
\end{figure*}

\begin{figure*}[!h]
  \centering
  \includegraphics[width=4.2in]{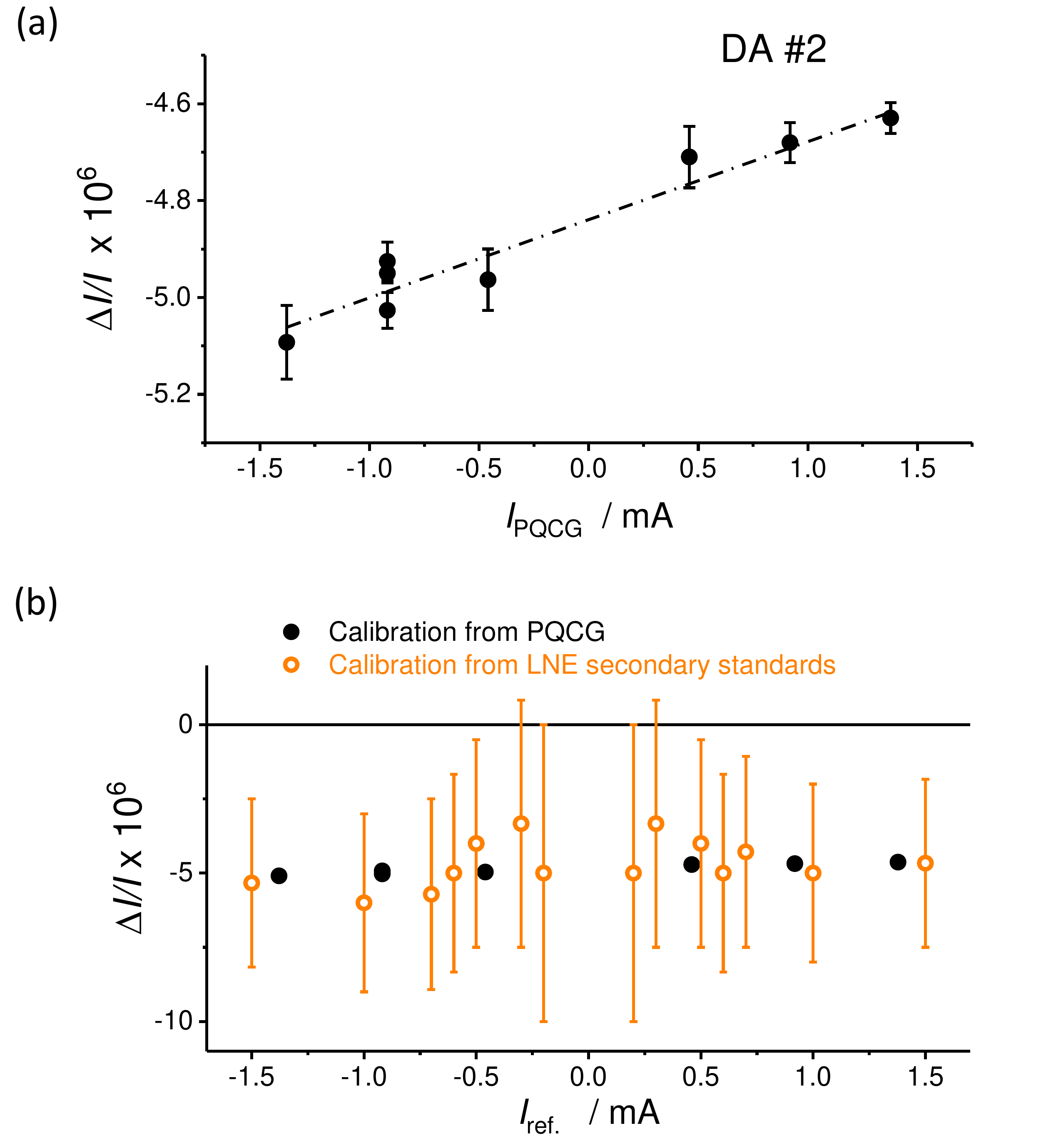}
  \caption{\small{(a) Calibration results on the 2-mA range of DA$\sharp$2. At 1.5 mA, $\Delta I/I = (I_\mathrm{meas.}-I_\mathrm{PQCG})/I_\mathrm{PQCG}$ is $-4.63\times10^{-6}$. A linear dependence on the current is resolved with a slope of $+1.6\times10^{-7}/$mA. Error bars are relative combined uncertainties ($k_\mathrm{c}=1$). (b) Comparison of the calibrations of the 2-mA range of the DA$\sharp$2 performed using the PQCG (black dots) and the LNE secondary standards (orange circles).} }
  \label{fig:Fig_Cal_Fluke}
\end{figure*}

The relative combined uncertainties of the calibrations are reported in figure\space\ref{fig:Fig_Cal_3458}, figure\space\ref{fig:Fig_Cal_3458-3} and figure\space\ref{fig:Fig_Cal_Fluke}. The Type B contribution due to the finite resolution of the DAs amounts to 3$\times10^{-8}$ at 1 mA. The combination of all other Type B contributions related to $I_\mathrm{PQCG}$, which are due to the CCC gain $G$, the Josephson frequency $f_\mathrm{J}$ and the correction $\alpha$, is estimated well below $10^{-8}$ \cite{Brun-Picard2016} and is therefore negligible.

Figure\space\ref{fig:Fig_Cal_3458}(a) reports the calibration results of DA$\sharp$1 performed using different settings of the PQCG. Results are expressed as the relative deviation $\Delta I/I = (I_\mathrm{meas.}-I_\mathrm{PQCG})/I_\mathrm{PQCG}$  between the measured current $I_\mathrm{meas.}$ and $I_\mathrm{PQCG}$, for two current levels: $\pm0.574$ mA and $\pm0.918$ mA. At 1 mA, the relative deviation $\Delta I/I$ is $-1.98\times10^{-6}$. The stability of $\Delta I/I$ over few hours is illustrated in the inset of figure \ref{fig:Fig_Cal_3458} from the repeatability of two sets of calibrations points. The dispersion between two identical measurements is of the order of $6\times10^{-7}$. A linear fit over all calibration points gives a very small slope of -0.7$\times10^{-7}/$mA which appears to be not significant with respect to the dispersion of the calibration points. This result indicates that the DA is linear within the stability of the instrument \cite{Spec3458-8508}.

During a calibration campaign, the quantization criteria demonstrated in reference \cite{Brun-Picard2016} have to be checked to ensure that the PQCG delivers quantized currents. They consist not only of the usual quantization criteria of the QHRS and of the PJVS but also of more specific criteria. The quantization of the generated current is tested: (i) by varying the PJVS bias current, (ii) by testing different combinations of Josephson array segments corresponding to the same number of Josephson junctions and (iii) by generating identical current levels with different numbers of junctions $n_\mathrm{J}$ and scaling the gain accordingly: for example, dividing $n_\mathrm{J}$ and multiplying the gain by the same factor should not change the measured current value. Figure\space\ref{fig:Fig_Cal_3458}(b) details the quantization criteria that have been tested during the calibration presented in figure\space\ref{fig:Fig_Cal_3458}(a). As is expected for a quantized current, the dispersion of the results obtained during the tests of the quantization criteria is of the order of the dispersion due to the instability of the DA observed during the calibration, and is much lower than the manufacturer’s specifications \cite{Spec3458-8508}.

Here, we would like to point out a notable difference with the previous results of reference \cite{Brun-Picard2016}, where the possibility to calibrate DAs on different ranges extending from $\mu$A to a few mA has been demonstrated. There, the tests of the quantization criteria highlighted an unexpected noise issue. It is illustrated in figure\space\ref{fig:Fig_Cal_3458-3}, which reports the relative combined uncertainties $u_c$ of the calibration points from Ref.\cite{Brun-Picard2016} (red squares) in the 1 mA range as a function of the number of Josephson junctions $n_\mathrm{J}$. Indeed, a clear dependence on $n_\mathrm{J}$ was observed, demonstrating that the noise was dominated by the PQCG rather than by the DA. This was rather unexpected as the relative Allan deviation expected for the PQCG alone, shown in figure\space\ref{fig:Fig_Cal_3458-2}b, is an order of magnitude below the relative Allan deviation of the DAs. We attributed this to the Johnson-Nyquist noise caused by the damping resistance $R_D$ at 293 K. 
The RC-filter formed by the 100-nF capacitance on the input resistance of the DA \cite{Brun-Picard2016} corresponds to a cutoff frequency of 16 kHz, which does not prevent the noise generated by the damping circuit from perturbing the system and overcoming the noise of the DA. The new uncertainties observed with $R_D$ at 4.2 K, reported as blue triangles in figure\space\ref{fig:Fig_Cal_3458-3}, are independent of $n_\mathrm{J}$, which confirms that the dominant contribution in the uncertainty now comes from the DA.

Figure\space\ref{fig:Fig_Cal_Fluke}(a) shows the results of the calibration of DA$\sharp$2 in the 2-mA range. $I_\mathrm{PQCG}$ has been generated with the gain $G = 160/4$ and with $n_\mathrm{J}$ = 3072, 2048 and 1024, corresponding to current levels $\pm$1.378 mA, $\pm$0.918 mA and $\pm$0.459 mA respectively. The dispersion of three calibration points (over 2 h) is about $1\times10^{-7}$ at -0.9 mA \cite{Spec3458-8508}. This allows the unveiling of a small but significant non linearity of DA$\sharp$2, revealed by a linear dependence of $\Delta I/I$ in the measured current range with a slope of $+1.6\times10^{-7}/$mA.

Figure\space\ref{fig:Fig_Cal_Fluke}(b) shows the comparison of the calibration of figure\space\ref{fig:Fig_Cal_Fluke}(a) with the calibration performed at LNE using secondary resistance and voltage standards with a relative uncertainty of ($5\times10^{-6}+1\times10^{-9}/I_\mathrm{meas.}$). Both calibrations are in good agreement within the measurement uncertainties despite the 6-months period separating them. This might indicate a good long-term stability of the DA, better than the manufacturer's specifications, which would have to be confirmed in a further study.

These calibration examples show that the PQCG is a precious tool which allows characterizing thoroughly precision DAs with relative uncertainties close to $10^{-7}$ while reducing the traceability chain. By doing regular calibrations with the PQCG, a better knowledge of the short and long term stabilities of such precision DAs can be gained. Hence, if regularly calibrated by the PQCG, they could be used as secondary standards with uncertainties better than the manufacturer specifications with the goal of improving the calibration and measurement capabilities (CMCs).

\section{ULCA comparison}\label{ULCA comp}
\subsection{Context}
In this last section, we present the results of the direct comparison of the current measurements performed using the Ultrastable Low-Noise Current Amplifier and the PQCG. The ULCA can be used for two purposes: current measurement or current generation, i.e., for current-to-voltage or voltage-to-current conversion. It is based on a two-stage design: a first stage provides a current gain $G_I$ of 1000 and a second stage provides a current-to-voltage conversion with a transresistance $R_{IV}$ (typically, 1  M$\Omega$ for the 'standard' ULCA). Thanks to this design, the device can be operated either in normal or extended mode \cite{Drung2017} which differ by the factor of 1000 in current range; for example, in the case of the 'standard' ULCA, the 5 nA range (normal mode) and the 5 $\mu$A range (extended mode) are accessible. Several versions of the ULCA have been developed \cite{Drung2017}. Among the second generation of ULCAs, the prototype of a high-accuracy variant ($R_{IV}$-nominal value 100 k$\Omega$) offers the highest current range compatible with the PQCG ($\pm50$ $\mu$A in the extended mode).

Thanks to its small size and compact design, the ULCA is perfectly suited as a travelling standard \cite{Drung2015,Giblin2019}. The use of an ULCA as a travelling standard has been exercised first in an interlaboratory comparison between three national metrology institutes including LNE and PTB in 2015 with a two-channel 'standard' ULCA \cite{Drung2015} operated in the 5 nA range. The calibration history of the traveling standard including the effects of lab-to-lab transports are crucial information; for instance, the history of the device which has been used in \cite{Drung2015} has been recorded over more than five years \cite{Krause2019}. However, comparably reliable information on the stability of the chosen high-accuracy ULCA, especially when travelling, was not available at the time of the first experiments reported below. The shipment from PTB to LNE was done by a commercial carrier and for the return transport, a PTB car has been used.

\subsection{Description of the comparison and results}
\begin{figure*}[!h]
  \centering
  \includegraphics[width=4.5in]{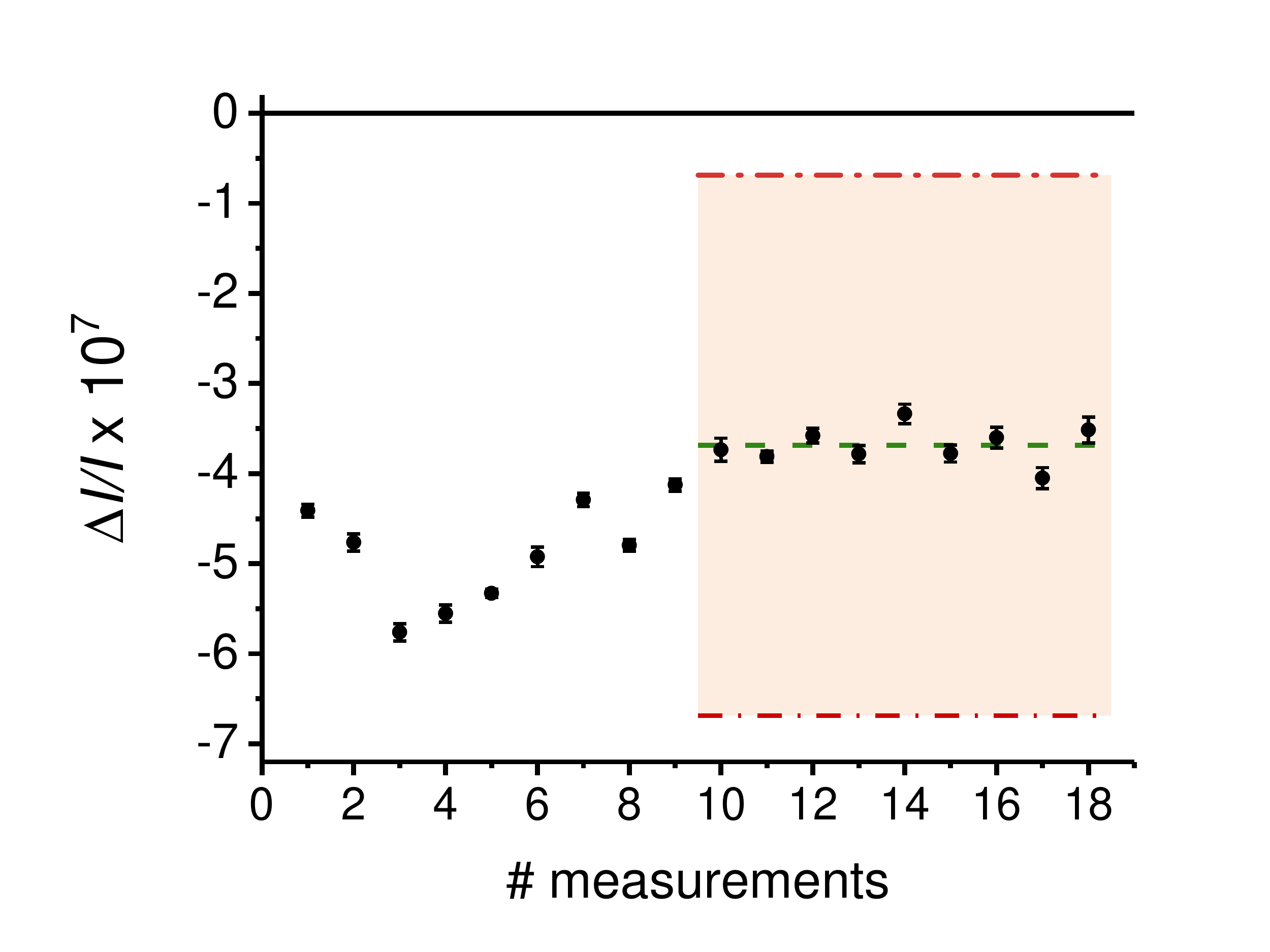}
  \caption{\small{Results of the comparison expressed as the relative deviation $\Delta I/I=(I_\mathrm{ULCA}-I_\mathrm{PQCG})/I_\mathrm{PQCG}$ (black dots). The small error bars represent the Type A standard uncertainty. The green dashed line at $-3.7\times10^{-7}$ represents the average of the last 9 points. The red dash dotted lines indicate the limits defined by the relative combined uncertainty $u_c = 3.1\times10^{-7}$ ($k_\mathrm{c}$ = 1)}.}
  \label{fig:comp_results}
\end{figure*}

The comparison has been done in extended mode in the 50 $\mu$A range at two current levels 43.75 $\mu$A and 21.875 $\mu$A. Similarly to the calibration of the DAs, $I_\mathrm{PQCG}$ is directly fed into the ULCA (i.e. into the current-to-voltage conversion stage). Different from the work with the DAs described in the previous section, no filter was necessary on the input of the ULCA. Introducing much less noise, the latter did not affect the SQUID’s stability. The calibration of the total transimpedance $A_{\mathrm{TR}} = G_I R_{IV}$, which is performed at PTB in two steps, has been done before and after travelling to LNE. The gain $G_I$ is directly calibrated by means of a CCC and the transresistance $R_{IV}$ is traced back to the quantum Hall resistance by means of a CCC-based resistance bridge \cite{Drung2015}. Note that for the present comparison with the ULCA operated in extended mode throughout, only the value of $R_{IV}$ is relevant.

\begin{table}[!h]
\newcolumntype{M}[1]{>{\centering\arraybackslash}m{#1}}
\begin{center}
\begin{tabular}{c c c c M{4cm} }
  \hline
  \textbf{$\pm$ $n_\mathrm{J}$} & \textbf{$N_\mathrm{JK}$} & \textbf{$N$} & \textbf{$\pm$ $I_\mathrm{PQCG}$} & \textbf{Tests}  \tabularnewline \hline \hline
  $\pm$ 2048&160&84&$\pm$ 43.750 $\mu$A& Nominal value \tabularnewline \hline
  $\pm$ 1024&160&84&$\pm$ 21.875 $\mu$A& Nominal value/2\tabularnewline \hline
  $\pm$ 2048&160&42&$\pm$ 43.750 $\mu$A& $n_\mathrm{J}/2$ and $G\times 2$ \tabularnewline \hline
  $\pm$ 2048&160&84&$\pm$ 43.750 $\mu$A& Bias trim: $\pm$0.1 mA and $\pm$0.2 mA\\  \hline
\end{tabular}
\caption{Parameters of the measurements during the comparison: number of Josephson junctions $n_\mathrm{J}$, number of turns $N_\mathrm{JK}$ and $N$, and tests realized.} \label{Comp50microAmp}
\label{tableau:mesure50microA}
\end{center}
\end{table}

The multiple connection of the PQCG has been implemented. However, to even further reduce the absolute value of the correction, we have 'short circuited' part of the cable resistances by using the “local triple connection” on one side of the QHRS (dashed connection on figure\space\ref{fig:schema}). That is, an additional connection is made between the voltage pad of the QHR and the top of the cryostat rather than directly on the superconducting pad of the PJVS. From the measurements of the resistances of the cables, we have estimated the relative correction $\alpha=1.35\times10^{-7}\pm2\times10^{-9}$.

\begin{figure*}[!h]
  \centering
  \includegraphics[width=4.5in]{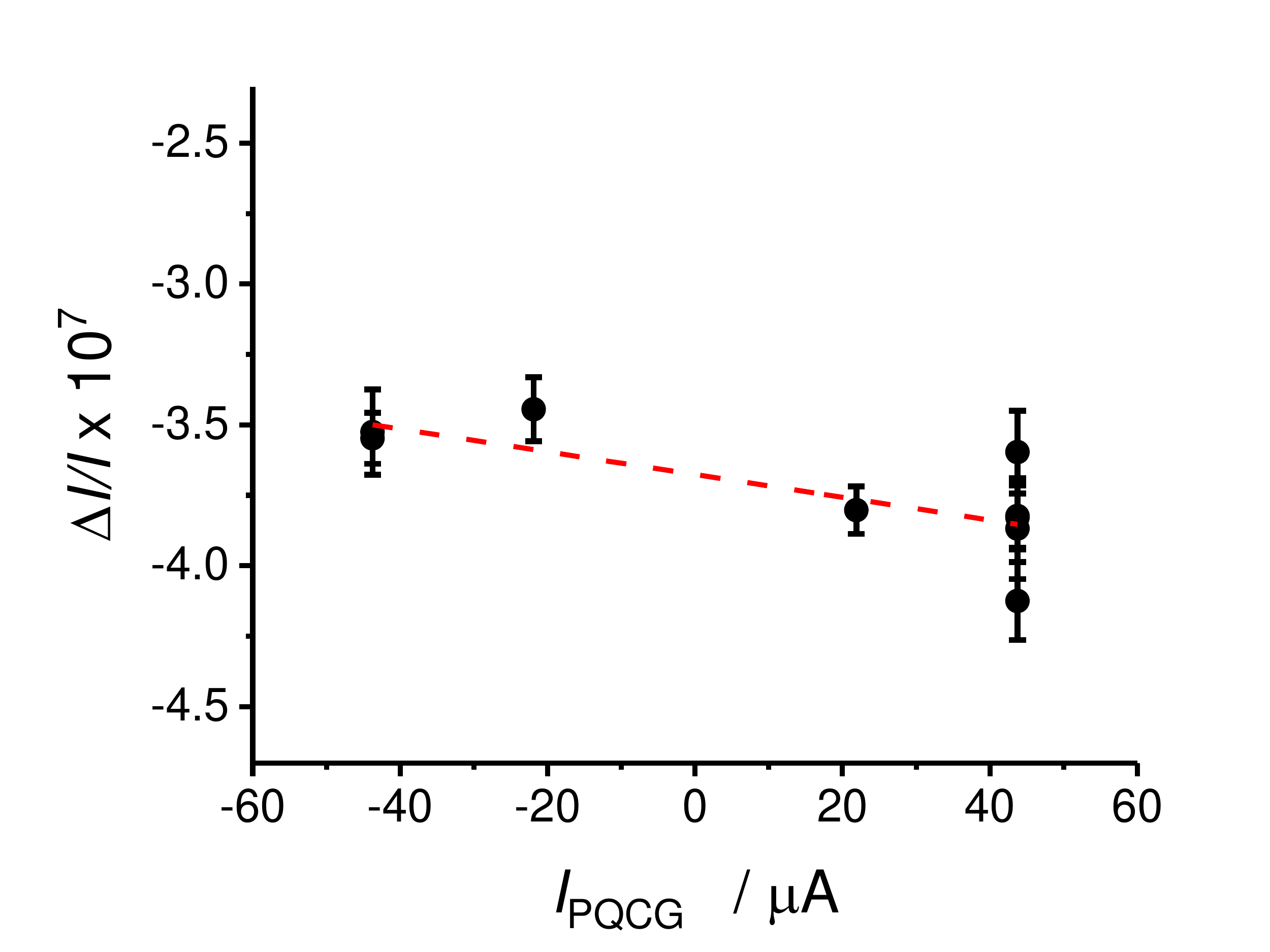}
  \caption{\small{Consistency check of the linearity of the ULCA on the $\pm$50-$\mu$A range done with the 9 last points of figure\space\ref{fig:comp_results}. The error bars represent the Type A standard uncertainty. }}
  \label{fig:ULCA_lineraity}
\end{figure*}

The current measured by the ULCA is given by $I_{\mathrm{ULCA}} = U_{R_{IV}}/R_{IV}$. It is determined from the value of the transresitance $R_{IV}$ estimated at the time of the comparison from the calibration history determined at PTB (see section \ref{RIV_at_PTB} in the annex to this paper) and from the voltage drop $U_{R_{IV}}$ across $R_{IV}$ measured on the 10 V range of a digitizing voltmeter (DVM). The latter is routinely calibrated against a 10-V conventional Josephson voltage standard at LNE.

In that way, we compare on one side the current measured by the combination of the ULCA and the DVM - both traceable to PTB’s quantum Hall resistor and LNE’s Josephson voltage standard, respectively - and the current generated on the other side by the PQCG - involving LNE’s PJVS and QHRS - and estimated from the expression $I_\mathrm{PQCG} = \frac{N_\mathrm{JK}}{N} \frac{U_\mathrm{J}}{R_\mathrm{H}}(1-\alpha)$. Considering the degrees of equivalence of PTB and LNE in the international comparison BIPM.EM-K10.b \cite{Solve2009} of Josephson array voltage standards at 10 V, this experiment can be interpreted as a comparison of the realizations of the ampere in the two laboratories. However, without the knowledge of the transresistance, this experiment can be used to determine the value $R_{IV}$ from $I_\mathrm{PQCG}$ and $U_{R_{IV}}$ (see section \ref{RIV_at_PTB}).

We have performed eighteen measurements within two days. We have applied exactly the same measurement sequence as for DA$\sharp1$. Figure\space\ref{fig:comp_results} shows the relative deviation $\Delta I/I= (I_\mathrm{ULCA}-I_\mathrm{PQCG})/I_\mathrm{PQCG}$ over the two days. These measurements include the tests of the quantization criteria of the PQCG and the consistency check in terms of linearity of the ULCA. A summary of the parameters of the measurements is presented in table \ref{tableau:mesure50microA}.

The Type A relative uncertainty of each measurement point is $<2\times10^{-8}$. This low Type A uncertainty allowed us to resolve a drift of 2.5$\times10^{-7}$ over the first 24 h followed by a period of stabilization to within 2.5$\times10^{-8}$, corresponding to the last 9 points in figure\space\ref{fig:comp_results}. The quantization criteria have been checked within the standard deviation of $\leq2.5\times10^{-8}$. After stabilization, the linearity of the ULCA has been tested at $\pm$ 21.875 $\mu$A and $\pm$ 43.750 $\mu$A. The results are shown in figure\space\ref{fig:ULCA_lineraity}. The error bars represent only the Type A standard uncertainty. A slight deviation of the order of $10^{-8}$ can be observed but it is within the stability of the measurements.

\begin{table}[!h]
\newcolumntype{M}[1]{>{\centering\arraybackslash}m{#1}}
\begin{center}
\begin{tabular}{c c c c }
  \hline
  \textbf{Contribution} & \textbf{$u_B$} \tabularnewline \hline \hline
  \small{ULCA: $R_{IV}$ calibration} &$6\times10^{-9}$ \tabularnewline \hline
  \small{ULCA: $R_{IV}$ temperature correction} & 2$\times10^{-8}$\tabularnewline \hline
  \small{ULCA: $R_{IV}$ stability} &$2\times10^{-7}$ \tabularnewline \hline
  \small{PQCG: cable correction}&$2\times10^{-9}$ \tabularnewline \hline
  \small{PQCG: other contributions} &$<2\times10^{-9}$ \tabularnewline \hline
  \small{DVM: calibration} & $2.3\times10^{-7}$ \tabularnewline \hline
  \small{\textbf{Total}} &$3.1 \times10^{-7}$  \\  \hline
\end{tabular}
\caption{Relative Type B standard uncertainty ($u_B$) budget of the comparison.}
\label{tableau:Table_Inc_budget}
\end{center}
\end{table}

The DVM gain error has been estimated from a two-weeks calibration campaign before the comparison. Unfortunately, we observed a deviation from the known linear drift when recalibrating the instrument one month after the comparison, which might have been due to a temperature regulation problem in the laboratory during this period.  Therefore the relative Type B uncertainty contribution due to this correction has been increased to cover this problem and amounts to 2.3$\times10^{-7}$.

The typical uncertainty of the calibrations of $R_{IV}$ at PTB is 6$\times10^{-9}$. During the comparison the internal temperature of the ULCA was recorded and was about 21 $^{\circ}$C at LNE, whereas the transresistance calibration is given at 23 $^{\circ}$C. This temperature difference was accounted for by a correction calculated from the $-1.4\times10^{-7}/ ^{\circ}$C temperature coefficient determined at PTB. The corresponding relative uncertainty contribution is 0.2$\times10^{-7}$. An additional uncertainty component has been added in the uncertainty budget, it accounts for the long term stability and the stability against transportation of $R_{IV}$. It has been estimated to 2$\times10^{-7}$ (see section \ref{RIV_at_PTB}). Table \ref{tableau:Table_Inc_budget} summarizes the Type B contributions to the relative standard uncertainty. The DVM calibration and the stability of the output stage of the ULCA with transportation appear as the dominant contributions.

For the final result of the comparison, we have used 9 data points obtained for $\pm$ 21.875\space$\mu$A and $\pm$ 43.750 $\mu$A. The experimental standard deviation of the mean of the 9 points is $7\times10^{-9}$. The final result of the comparison is expressed as the mean value of the 9 points within the total relative combined uncertainty (represented by the red dashed dotted lines in figure\space\ref{fig:comp_results})($k_\mathrm{c}=1$):
\begin{equation}\label{final result}
  [I_\mathrm{ULCA}-I_\mathrm{PQCG}]/I_\mathrm{PQCG} = (-3.7\pm 3.1)\times10^{-7}.
\end{equation}
In this comparison, the realizations of the ampere from both laboratories agree to -3.7 parts in $10^{7}$ with a combined standard uncertainty of 3.1 parts in $10^{7}$ ($k_\mathrm{c}=1$). This result is reported in Figure \ref{fig:comp_results}. Clearly there is no significant discrepancy between the two current values since the difference is lower than the expanded uncertainty of 6.2 parts in $10^{7}$ ($k_\mathrm{c}=2$).

\section{Conclusion}

To conclude, we have identified and estimated the noise sources of the PQCG with the support of an electrical model based on magnetically coupled RLC circuits. It turns out that the primary quantum circuit of the PQCG, which is formed by the PJVS, the QHRS and the CCC in series, is strongly coupled to noise via the connection to ground. This leads to an instability of the SQUID for windings of large number of turns, which is resolved by implementing a cryogenic damping circuit, an optimized SQUID feedback circuit and an adapted grounding scheme.

These improvements allowed us performing calibrations of commercial precision ammeters with a relative measurement uncertainty of a few parts in $10^{7}$, only limited by their stability. These devices, which also demonstrated linearity within a few parts in $10^{7}$, could be used as secondary current standards if they are periodically calibrated against the PQCG.

Furthermore, the improved PQCG was used to feed an Ultrastable Low-Noise Current Amplifier (ULCA) with currents in the 50 $\mu$A range. The agreement we found, at a level of a few parts in $10^{7}$, demonstrates the equivalence of the ampere realizations at LNE and PTB. Our results represent the state-of-the-art of currents in the range $\pm$ 50 $\mu$A. However, there is room for improvements for future comparisons with the goal to reach 1 part in $10^{7}$ or less. Improvements should include: (i) the use of transportable thermostat (thus avoiding thermal shocks during transport potentially responsible for discontinuity in the calibration value of the ULCA), (ii) timely calibration of the DVM before and after the series of measurements (or even during this series). Even better performance could be achieved by using a null detection scheme with a second PJVS \cite{Brun-Picard2016}. Moreover, the range over which the currents can be compared can be extended down to tens of the nA \cite{Djordjevic_2020} with the same expected uncertainties. All these new results open the way to a future metrology of the ampere that should be as efficient as those of the ohm and the volt.

\clearpage

\section{Acknowledgments}
The authors gratefully acknowledge support by M. Mghalfi, J. Azib and O. N. Ndir. This work was supported by the Joint Research Project ‘e-SI-Amp’ (15SIB08). It received funding from the European Metrology Programme for Innovation and Research (EMPIR) cofinanced by the Participating States and from the European Union’s Horizon 2020 research and innovation programme.


\section{References}

\begin{thebibliography}{10}
\expandafter\ifx\csname url\endcsname\relax
  \def\url#1{\texttt{#1}}\fi
\expandafter\ifx\csname urlprefix\endcsname\relax\def\urlprefix{URL }\fi
\expandafter\ifx\csname href\endcsname\relax
  \def\href#1#2{#2} \def\path#1{#1}\fi

\bibitem{Brun-Picard2016}
J.~Brun-Picard, S.~Djordjevic, D.~Leprat, F.~Schopfer, W.~Poirier, Practical
  quantum realization of the ampere from the elementary charge, Phys. Rev. X 6
  (2016) 041051.

\bibitem{SI_Brochure_2019}
BIPM, The {I}nternational {S}ystem of {U}nits ({SI}), 9th Edition, Bureau
  international des poids et mesures, Pavillon de Breteuil, {F}-92312
  {S}\`{e}vres {C}edex, France, 2019.

\bibitem{Poirier2019}
W.~Poirier, S.~Djordjevic, F.~Schopfer, O.~Th\'evenot, The ampere and the
  electrical units in the quantum era, C. R. Physique 20 (2019) 92.

\bibitem{Pekola2013}
J.~P. Pekola, O.~P. Saira, V.~Maisi, A.~Kemppinen,
  M.~M$\mathrm{\ddot{o}}$tt$\mathrm{\ddot{o}}$nen, Y.~A. Pashkin, D.~Averin,
  Single-electron current sources: Toward a refined definition of the ampere,
  Rev. Mod. Phys. 85 (2013) 1421--1472.

\bibitem{Scherer2019}
H.~J. Scherer, H.~W. Schumacher,
  \href{https://onlinelibrary.wiley.com/doi/abs/10.1002/andp.201800371}{Single-electron
  pumps and quantum current metrology in the revised {SI}}, Annalen der Physik
  531~(5)  1800371.
\newblock \href
  {http://arxiv.org/abs/https://onlinelibrary.wiley.com/doi/pdf/10.1002/andp.201800371}
  {\path{arXiv:https://onlinelibrary.wiley.com/doi/pdf/10.1002/andp.201800371}},
  \href {http://dx.doi.org/10.1002/andp.201800371}
  {\path{doi:10.1002/andp.201800371}}.
\newline\urlprefix\url{https://onlinelibrary.wiley.com/doi/abs/10.1002/andp.201800371}

\bibitem{Giblin_2019}
S.~P. Giblin, A.~Fujiwara, G.~Yamahata, M.-H. Bae, N.~Kim, A.~Rossi,
  M.~M\"{o}tt\"{o}nen, M.~Kataoka,
  \href{https://doi.org/10.1088%2F1681-7575%2Fab29a5}{Evidence for universality
  of tunable-barrier electron pumps}, Metrologia 56~(4) (2019) 044004.
\newblock \href {http://dx.doi.org/10.1088/1681-7575/ab29a5}
  {\path{doi:10.1088/1681-7575/ab29a5}}.
\newline\urlprefix\url{https://doi.org/10.1088%2F1681-7575%2Fab29a5}

\bibitem{Kaneko_2016}
N.-H. Kaneko, S.~Nakamura, Y.~Okazaki,
  \href{https://doi.org/10.1088/0957-0233/27/3/032001}{A review of the quantum
  current standard}, Measurement Science and Technology 27~(3) (2016) 032001.
\newblock \href {http://dx.doi.org/10.1088/0957-0233/27/3/032001}
  {\path{doi:10.1088/0957-0233/27/3/032001}}.
\newline\urlprefix\url{https://doi.org/10.1088/0957-0233/27/3/032001}

\bibitem{Stein2017}
F.~Stein, H.~Scherer, T.~Gerster, R.~Behr, M.~G{\"{o}}tz, E.~Pesel, C.~Leicht,
  N.~Ubbelohde, T.~Weimann, K.~Pierz, H.~W. Schumacher, F.~Hohls, Robustness of
  single-electron pumps at sub-ppm current accuracy level, Metrologia 54 (2017)
  S1.

\bibitem{Chae_2020}
D.-H. Chae, M.-S. Kim, W.-S. Kim, T.~Oe, N.-H. Kaneko,
  \href{https://doi.org/10.1088%2F1681-7575%2Fab605f}{Quantum mechanical
  current-to-voltage conversion with quantum hall resistance array}, Metrologia
  57~(2) (2020) 025004.
\newblock \href {http://dx.doi.org/10.1088/1681-7575/ab605f}
  {\path{doi:10.1088/1681-7575/ab605f}}.
\newline\urlprefix\url{https://doi.org/10.1088%2F1681-7575%2Fab605f}

\bibitem{Fan_2019}
I.~{Fan}, R.~{Behr}, D.~{Drung}, C.~{Krause}, M.~{G\"{o}tz}, E.~{Pesel},
  H.~{Scherer}, Externally referenced current source with stability down to 1
  n{A}/{A} at 50 m{A}, IEEE Transactions on Instrumentation and Measurement
  68~(6) (2019) 2129--2135.
\newblock \href {http://dx.doi.org/10.1109/TIM.2018.2885502}
  {\path{doi:10.1109/TIM.2018.2885502}}.

\bibitem{DrungRSI2015}
D.~Drung, C.~Krause, U.~Becker, H.~Scherer, F.~Ahlers, Ultrasatable low-noise
  current amplifier: A novel device for measuring small electric currents with
  high accuracy, Rev. Sci. Instrum. 86 (2015) 024703.

\bibitem{Drung2017}
D.~Drung, C.~Krause, Ultrastable low-noise current amplifiers with extended
  range and improved accuracy, IEEE Trans. Intrum. Meas. 66 (2017) 1425.

\bibitem{Delahaye1993}
F.~Delahaye, Series and parallel connection of multiterminal quantum
  hall-effect devices, J. Appl. Phys. 73 (1993) 7914.

\bibitem{Poirier2014}
W.~Poirier, F.~Lafont, S.~Djordjevic, F.~Schopfer, L.~Devoille, A programmable
  quantum current standard from the josephson and the quantum hall effects, J.
  Appl. Phys. 115 (2014) 044509.

\bibitem{Poirier2020}
W.~Poirier, D.~Leprat, F.~Schopfer, A resistance bridge based on a cryogenic
  current comparator achieving sub-$10^{-9}$ measurement uncertainties, IEEE
  Trans. Instrum. Meas. 70 (2021) 1.

\bibitem{Mueller2007}
F.~Mueller, R.~Behr, L.~Palafox, J.~Kohlmann, R.~Wendisch, I.~Krasnopolin,
  Improved 10 {V} {SINIS} series arrays for applications in {AC} voltage
  metrology, IEEE Trans. Appl. Supercond. 17 (2007) 649.

\bibitem{Muller2009}
F.~M{\"{u}}ller, R.~Behr, T.~Weimann, L.~Palafox, D.~Olaya, P.~D. Dresselhaus,
  S.~P. Benz, 1 {V} and 10 {V} {SNS} programmable voltage standards for 70
  {GH}z, IEEE Transactions on Applied Superconductivity 19~(3) (2009) 981--986.
\newblock \href {http://dx.doi.org/10.1109/TASC.2009.2017911}
  {\path{doi:10.1109/TASC.2009.2017911}}.

\bibitem{Piquemal1993}
F.~Piquemal, G.~Genev\`es, F.~Delahaye, J.~P. Andr\'e, J.~N. Patillon,
  P.~Frijlink, Report on a joint {BIPM-EUROMET} project for the fabrication of
  {QHE} samples by the {LEP}s, IEEE Trans. Instrum. Meas. 42 (1993) 264.

\bibitem{Drung2009}
D.~Drung, M.~G{\"{o}}tz, E.~Pesel, J.-H. Storm, C.~Assmann, M.~Peters,
  T.~Schurig, Improving the stability of cryogenic current comparator setups,
  Supercond. Sci. Technol. 22 (2009) 114004.

\bibitem{Azib2018}
J.~Azib, J.~Brun-Picard, F.~Schopfer, W.~Poirier, S.~Djordjevic, Towards an
  improved programmable quantum current generator, 2018 Conference on Precision
  Electromagnetic Measurements (CPEM 2018) (2018) 1--2.

\bibitem{Rengnez_2015}
F.~Rengnez, D\'{e}veloppement de comparateur cryog\'{e}nique de courants
  tr\`{e}s faible bruit pour la m\'{e}trologie \'{e}lectrique quantique., Ph.D.
  thesis, Universit\'{e} Paris-Saclay (2015).

\bibitem{Remark1}
One could similarly consider an inductive coupling of the circuit with an
  external source).

\bibitem{Williams2020}
J.~M. Williams, J.~Ireland, I.~Rungger,
  \href{https://doi.org/10.1063/5.0019704}{Application of conserved
  thermal-noise energy model to damped resonant behavior in cryogenic current
  comparators}, Review of Scientific Instruments 91~(10) (2020) 105115.
\newblock \href {http://dx.doi.org/10.1063/5.0019704}
  {\path{doi:10.1063/5.0019704}}.
\newline\urlprefix\url{https://doi.org/10.1063/5.0019704}

\bibitem{Drung1996}
D.~Drung, \textsc{SQUID} Sensors: Fundamentals, Fabrication and Applications,
  Ed. H. Weinstock, Springer, Netherlands, 1996, Ch. Advanced \textsc{SQUID}
  read-out electronics, pp. 63--116.

\bibitem{SQUID_QuantumDesign_550}
Quantum Design, Manual of Controller 550.

\bibitem{Sanchez2009}
C.~A. Sanchez, B.~M. Wood, A.~D. Inglis, {CCC} bridge with digitally controlled
  current sources, IEEE Trans. Instrum. Meas. 58 (2009) 1202.

\bibitem{Williams2011}
J.~M. {Williams}, G.~{Rietveld}, E.~{Houtzager}, T.~J. B.~M. {Janssen}, Design
  considerations for a {CCC} bridge with complete digital control, IEEE
  Transactions on Instrumentation and Measurement 60~(12) (2011) 3907--3912.
\newblock \href {http://dx.doi.org/10.1109/TIM.2011.2149330}
  {\path{doi:10.1109/TIM.2011.2149330}}.

\bibitem{Drung2004}
D.~Drung, M.~M\"{u}ck, \textsc{SQUID} Handbook vol. I, Ed. J Clarke and A.
  Braginski, WILEY-VCH Verlag GmbH and Co. KGaA, Germany, 2004, Ch.
  \textsc{SQUID} electronics, pp. 127--170.

\bibitem{1_over_f_noise}
Some spectra showed $1/f$ noise below 1 {H}z, but for the moment, we lack of
  reliable data in this frequency range. {A} rough estimate of $1/f$ noise by
  linear extrapolation could amount to about 20 to 50
  $\mu\phi_{0}$/{H}z$^{1/2}$ at 20 m{H}z.

\bibitem{Witt2005}
T.~Witt, Allan variances and spectral densities for dc voltage measurements
  with polarity reversals, IEEE Trans. Instrum. Meas. 54 (2005) 550.

\bibitem{Spec3458-8508}
From the constructors ,we use the "accuracy after 24 h" ({DA}$\sharp1$) and the
  "uncertainty relative to calibration standards after 24 hours"
  ({DA}$\sharp2$) to estimate the stability of the instruments over 24 h (it
  includes a contribution from the noise of the instrument). {A}t 1 m{A}, in
  the case of {DA}$\sharp1$, the accuracy after 24 h (for {NPLC} 100) is ± (10
  ppm {R}eading + 4 ppm {R}ange), corresponding to a relative unceratinty of
  14$\times10^{-6}$ ($k_c$=2); for {DA}$\sharp2$, it is equal to ± (5 ppm
  {R}eading + 2 ppm {R}ange) corresponding to a relative unceratinty of
  $9.5\times10^{-6}$ ($k_c$=2).

\bibitem{Drung2015}
D.~Drung, C.~Krause, S.~P. Giblin, S.~Djordjevic, F.~Piquemal, O.~S{\'{e}}ron,
  F.~Rengnez, M.~G{\"{o}}tz, E.~Pesel, H.~Scherer,
  \href{https://doi.org/10.1088/0026-1394/52/6/756}{Validation of the
  ultrastable low-noise current amplifier as travelling standard for small
  direct currents}, Metrologia 52~(6) (2015) 756--763.
\newblock \href {http://dx.doi.org/10.1088/0026-1394/52/6/756}
  {\path{doi:10.1088/0026-1394/52/6/756}}.
\newline\urlprefix\url{https://doi.org/10.1088/0026-1394/52/6/756}

\bibitem{Giblin2019}
S.~P. {Giblin}, D.~{Drung}, M.~{G\"{o}tz}, H.~{Scherer}, Interlaboratory
  nanoamp current comparison with subpart-per-million uncertainty, IEEE
  Transactions on Instrumentation and Measurement 68~(6) (2019) 1996--2002.
\newblock \href {http://dx.doi.org/10.1109/TIM.2018.2879126}
  {\path{doi:10.1109/TIM.2018.2879126}}.

\bibitem{Krause2019}
C.~Krause, D.~Drung, M.~G{\"{o}}tz, H.~Scherer,
  \href{https://doi.org/10.1063/1.5078572}{Noise-optimized ultrastable
  low-noise current amplifier}, Review of Scientific Instruments 90~(1) (2019)
  014706.
\newblock \href {http://arxiv.org/abs/https://doi.org/10.1063/1.5078572}
  {\path{arXiv:https://doi.org/10.1063/1.5078572}}, \href
  {http://dx.doi.org/10.1063/1.5078572} {\path{doi:10.1063/1.5078572}}.
\newline\urlprefix\url{https://doi.org/10.1063/1.5078572}

\bibitem{Solve2009}
S.~Solve, R.~Chayramy, S.~Djorjevic, O.~S\'eron,
  \href{http://stacks.iop.org/0026-1394/46/i=1A/a=01002}{Comparison of the
  {J}osephson voltage standards of the {LNE} and the {BIPM} (part of the
  ongoing bipm key comparison {BIPM.EM-K}10.b)}, Metrologia 46~(1A) (2009)
  01002.
\newline\urlprefix\url{http://stacks.iop.org/0026-1394/46/i=1A/a=01002}

\bibitem{Djordjevic_2020}
S.~{Djordjevic}, W.~{Poirier}, D.~{Drung}, M.~{G\"{o}tz}, Comparison of the
  programmable quantum current generator and an ultrastable low-noise current
  amplifier, in: 2020 Conference on Precision Electromagnetic Measurements
  (CPEM), 2020, pp. 1--2.
\newblock \href {http://dx.doi.org/10.1109/CPEM49742.2020.9191863}
  {\path{doi:10.1109/CPEM49742.2020.9191863}}.

\end{thebibliography}
\bibliographystyle{elsarticle-num}

\section{Annex}

\subsection{PJVS, QHRS and CCC details}\label{exp_details}
The Josephson array is subdivided into 14 smaller array segments, and follows the sequence 256/512/3072/2048/1024/128/1/1/2/4/8/16/32/64. The quantum Hall resistance standard, based on an eight-terminal Hall bar made of a GaAs/AlGaAs semiconductor heterostructure (LEP514), was produced at the Laboratoire Electronique de Philips \cite{Piquemal1993}. 
The design of the CCC used here is detailed in \cite{Poirier2020}. It is composed of 15 windings with the following numbers of turns: 1, 1, 2, 2, 16, 16, 32, 64, 128, 160, 160, 1600, 1600, 2065, and 2065. It is equipped with a Quantum Design Inc. dc SQUID, operating with flux modulation at 500 kHz. The SQUID has a white noise level of 3 $\mu\phi_{0}/\sqrt{\mathrm{Hz}}$ and a $1/f$ corner frequency $f_{c}$ = 0.3 Hz, where $\phi_{0} = K_\mathrm{J}^{-1}$ is the superconducting flux quantum. The flux to ampere-turn sensitivity of the CCC is $\gamma_{CCC}$ = 8 $\mu$A$\cdot$turn/$\phi_{0}$.  Note that in the present study no additional current divider \cite{Brun-Picard2016} is used to finely adjust the gain between $I_\mathrm{PQCG}$ and $I_\mathrm{PQCS}$.

\subsection{Model used for the simulations}\label{model}

\begin{figure*}[!h]
  \centering
  \includegraphics[width=3.2in]{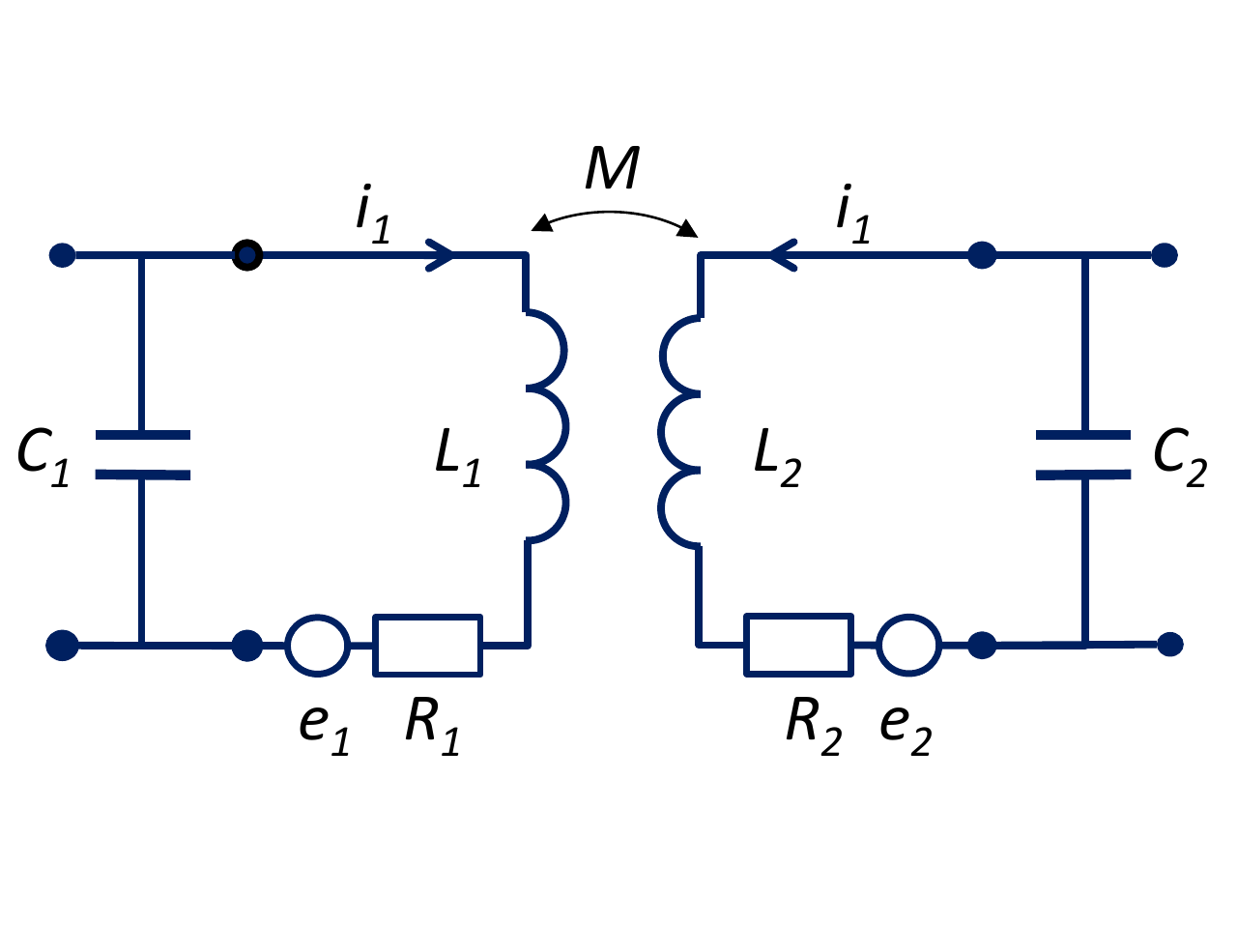}
  \caption{\small{Model of two coupled series-RLC circuits}}
  \label{fig:model_1}
\end{figure*}

For the model of the noise coming from the CCC at high frequency, we consider two series-RLC circuits ($R_1, C_1, L_1, R_2, C_2, L_2$) coupled by a mutual inductance $M=\kappa\sqrt{L_1L_2}$, where $\kappa$ is the coupling factor. The parameters are given in the figure\space\ref{fig:model_1}. The currents circulating in the two circuits $i_1$ and $i_2$ are written as a current vector:

\begin{equation*}
  I = \binom{i_1}{i_2}
\end{equation*}

The noise voltage density $\tilde{e}_1$ and $\tilde{e}_2$ of the voltage noise sources, $e_1$ and $e_2$, due to the Johnson-Nyquist noise of the resistances $R_1$ and $R_2$ are expressed in V/$\sqrt{\mathrm{Hz}}$ and are given by : ${\tilde{e}_1}=\sqrt{4kTR_1}$ and ${\tilde{e}_2}=\sqrt{4kTR_2}$ respectively and are written as a vector :
\begin{equation*}
  \tilde{E} = \binom{\tilde{e_1}}{\tilde{e_2}}
\end{equation*}
 The currents are deduced from the following system of linear equations :
\begin{equation*}
   I = [K]^{-1} \tilde{E}
\end{equation*}

where

\begin{equation*}
  K =
  \begin{pmatrix}
    j\frac{L_1}{\omega}(\omega^{2}-j\gamma_{1}\omega-\omega_{01}^{2})&j M\omega\\
    j M\omega &j \frac{L_2}{\omega}(\omega^{2}-j\gamma_{2}\omega-\omega_{02}^{2})
  \end{pmatrix}
\end{equation*}

and $\gamma_{1} = R_1/L_1, \gamma_{2} = R_2/L_2, \omega_{01} = \sqrt{\frac{1}{L_1C_1}}$ and $\omega_{02} = \sqrt{\frac{1}{L_2C_2}}$.

The screening current $\delta I$ on the toroidal superconducting shield due to the noise currents is deduced from the relation: $\delta I = N_1i_1 +N_2i_2 = \eta \tilde{e}_1 + \beta \tilde{e}_2$, where $\eta=N_1 K^{-1}_{11}+N_2 K^{-1}_{12}$ and $\beta=N_1K^{-1}_{12}+N_2K^{-1}_{22}$. The noise sources $e_1$ and $e_2$ being uncorrelated, the current noise amplitude is given by : $\sqrt{|\delta I|^2} = \sqrt{|\eta|^2 \tilde{e}_1^2+|\beta|^2 \tilde{e}_2^2}$. Finally, the flux noise contribution is $\sqrt{S_{CCC}(f)}=\sqrt{|\delta I|^2}/\gamma_{CCC}$.

\subsection{Calculation of the expected relative Allan deviation for the PQCG in figure\space 7}\label{Calc_Avar}
The Allan deviation allows the analysis of the noise by differentiating noise types according to the exponent of its power dependence with time. White noise manifests itself by a $\tau^{-1/2}$ dependence. In the white noise regime, the Allan deviation $\sigma_w$ is related to the flux power spectral density $S_\phi$ by :
\begin{equation}\label{Allan dev}
  \sigma_w = \sqrt{S_\phi/2\tau}
\end{equation}
where $\tau$ is the sampling time.

We want to relate the flux power spectral density measured for the PQCG to the relative Allan deviation $\sigma_I/I$ ($= \sigma_\phi/\phi$) expected for a series of current measurements obtained from on-off-on cycles that allows us to subtract the offsets. For that, we calculate the standard deviation of an on-off-on cycle by adapting a calculation done for polarity reversals by Drung et \emph{al.} \cite{DrungRSI2015}.

The value of the current calibration point, corresponding to one on-off-on cycle, is calculated from the average of the 'on' values for which we subtract the average of the 'off' values. We suppose that the standard deviations, $\sigma_{on}$ and $\sigma_{off}$, for the 'on' and 'off' polarities, can be calculated from the white noise standard deviations (see equation \ref{Allan dev}) corresponding to the duration, $\tau_{on}$ and $\tau_{off}=2\tau_{on}$, of the 'on' and 'off' polarities, respectively.

The white noise uncertainty for the on-off-on cycle, $\sigma_{cycle}$, is then given by :
\begin{equation}\label{standard dev}
  \sigma_{cycle}=\sqrt{\sigma_{on}^2/2+\sigma_{off}^2}= \sigma_{on}= \sqrt{S_\phi/2\tau_{on}}
\end{equation}
with $\sigma_{off}^2 = \sigma_{on}^2/2$.

In figure \ref{fig:Fig_Cal_3458-2}, we plot the relative Allan deviation as a function of the sampling time $\tau = n_s \tau_0$, so we want to express $\sigma_{cycle}$ as a function of $\tau_0$.
Following \cite{DrungRSI2015}, we define the effective measuring time $\tau_e$, i.e. the sampling time without the waiting time after polarity reversals, here: $\tau_{e}=\tau_0-2\tau_w=\tau_0(1-\rho)$, with $\rho=2\tau_w/\tau_0$ the rejection rate. For the on-off-on cycle, $\tau_{on}=\tau_{e}/4$. We rewrite $\sigma_{cycle}$ as a function of $\tau_0$,
\begin{equation}\label{standard dev}
  \sigma_{cycle}= \sqrt{2S_\phi/\tau_{e}} = \sqrt{2S_\phi/\tau_0(1-\rho)}
\end{equation}

We introduce the factor $A'$ defined as $A' = \sqrt{2S_\phi/(1-\rho)}$, such that :
\begin{equation}\label{standard dev}
  \sigma_{cycle}=  A'/\sqrt{\tau_0}
\end{equation}
Finally, the Allan deviation for a sequence of $n_s$ cycles is :
\begin{equation}\label{stand dev tau0}
  \sigma_{sequence} = A'/\sqrt{n_s\tau_{0}}
\end{equation}

All the parameters used for the calibration measurements and for the Allan deviation are given in table \ref{Parameters_data_acquisition}. We tried to realize as much as possible similar acquisition times for the two DAs despite the different parameter settings and the different dead times. As can be seen from table \ref{Parameters_data_acquisition}, the DA$\sharp1$ has been used in the mode AUTOZERO ON (internal offset errors due to bias currents of the amplifiers and thermal emfs are subtracted), which is the most accurate mode. However compared to the other DA, which does not have this function, the measuring time has been doubled for DA$\sharp1$, in order to have the same effective measuring time in a given polarity. This explains the difference of the timing parameters for the two DAs.

\begin{table}[!h]
\newcolumntype{M}[1]{>{\centering\arraybackslash}m{#1}}
\begin{center}
\begin{tabular}{c c c c M{4cm} }
  \hline
  \textbf{\small{Parameters}} & \textbf{\small{DA$\sharp$1 cal.}} & \textbf{\small{DA$\sharp$2 cal.}} & \textbf{\small{DA$\sharp$1 'Allan'}} & \textbf{\small{DA$\sharp$2 'Allan'}}  \tabularnewline \hline \hline
  \small{NPLC}&\small{10}&\small{64}&\small{10}&\small{64} \tabularnewline \hline
  \small{AUTOZERO}&\small{ON}&\small{/}&\small{ON}&\small{/} \tabularnewline \hline
  \small{Points per pol.}&\small{30}&\small{5}&\small{15}&\small{3} \tabularnewline \hline
  \small{$\tau_{on}$ (s)}&\small{12}&\small{6.4}&\small{6}&\small{3.8}\tabularnewline \hline
  \small{$\tau_{w}$ (s)}&\small{2}&\small{5}&\small{2}&\small{5}\tabularnewline \hline
  \small{$\tau_{0}$ (s)}&\small{52}&\small{35.6}&\small{28}&\small{25.4}\tabularnewline \hline
  \small{$f_{acq.}=1/\tau_{0}$ (mHz)}&\small{19}&\small{28}&\small{36}&\small{39}\tabularnewline \hline
  \small{$\tau_{e}$}&\small{48}&\small{25.6}&\small{24}&\small{15.4}\tabularnewline \hline
  \small{$\rho$}&\small{0.08}&\small{0.28}&\small{0.14}&\small{0.39}\tabularnewline \hline
  \small{$\sharp$ of points}&\small{11}&\small{11}&\small{71}&\small{71}\tabularnewline \hline
  \small{Total duration (s)}&\small{572}&\small{392}&\small{1988}&\small{1800}\\  \hline
  \label{Parameters_data_acquisition}
\end{tabular}
\caption{Parameters used for the calibration measurements and for the Allan deviation measurements.}
\end{center}
\end{table}

To calculate the relative Allan deviation expected for the PQCG at $I_{\mathrm{PQCG}}$ = 0.918 mA, we have to divide $A'$ by the flux seen by the SQUID due to $I_{\mathrm{PQCS}}$, i.e. $A'_{rel}$. The flux is 460 $\phi_0$ (calculated with the parameters $N_{\mathrm{JK}}$ = 160, $N$ = 4, $I_{\mathrm{PQCS}}$ = 23 $\mu$A and the flux sensitivity 8 $\mu$A$\cdot$turns/$\phi_0$). We calculate $A'_{rel}$ based on a large range of estimates for $\sqrt{S_\phi}$ at 20 mHz, which spans from 8 to 50 $\mu\phi_0$/$\sqrt{\mathrm{Hz}}$ taking into account the all cases from white noise to $1/f$ noise.

Based on the previous calculation, it is possible to use the fitting parameters $A'_{rel}$ of the relative Allan deviation $\sigma_I/I$ of the current measurements of the 2 DAs presented in figure \ref{fig:Fig_Cal_3458-2}, to estimate the equivalent current noise $\sqrt{S_I} = A'_{rel}\sqrt{(1-\rho)/2}*I_{\mathrm{PQCG}}$ and eventually to compare with the constructor specifications.

\begin{table}[!h]
\newcolumntype{M}[1]{>{\centering\arraybackslash}m{#1}}
\begin{center}
\begin{tabular}{c c c c }
  \hline
  \textbf{\small{Instrument}} & \textbf{\small{$A'_{rel}$}} & \textbf{\small{$\sqrt{S_{\phi}}$}} & \textbf{\small{$\sqrt{S_{I}}$}} \tabularnewline \hline \hline
  \small{PQCG}&\small{$2.6\times10^{-8}/\sqrt{\mathrm{Hz}}$ (*)}&\small{8 $\mu\phi_0$/$\sqrt{\mathrm{Hz}}$}&\small{/} \tabularnewline \hline
  \small{PQCG}&\small{$1.6\times10^{-7}/\sqrt{\mathrm{Hz}}$ (*)}&\small{50 $\mu\phi_0$/$\sqrt{\mathrm{Hz}}$}&\small{/} \tabularnewline \hline
  \small{DA$\sharp$1}&\small{$1.35\times10^{-6}/\sqrt{\mathrm{Hz}}$ (**)}&\small{/}&\small{0.8 nA/$\sqrt{\mathrm{Hz}}$(*)} \tabularnewline \hline
  \small{DA$\sharp$2}&\small{$6\times10^{-7}/\sqrt{\mathrm{Hz}}$ (**)}&\small{/}&\small{0.3 nA/$\sqrt{\mathrm{Hz}}$(*)}\\  \hline
\end{tabular}
\caption{Results of the calculation of $A'_{rel}$ for the PQCG when $\sqrt{S_\phi}$ at 20 mHz is equal to 8 or 50 $\mu\phi_0$/$\sqrt{\mathrm{Hz}}$ and results of the calculation of $\sqrt{S_I}$ for DA$\sharp1$ and DA$\sharp2$ from the values of $A'_{rel}$ obtained from the fit of figure \ref{fig:Fig_Cal_3458-2}. (*) denotes a result from a caculation, (**) denotes a result from a fit.}
\label{Calculation_figure_Allan}
\end{center}
\end{table}

\subsection{Determination of the $R_{IV}$}\label{RIV_at_PTB}
\begin{figure*}[!h]
  \centering
  \includegraphics[width=4.5in]{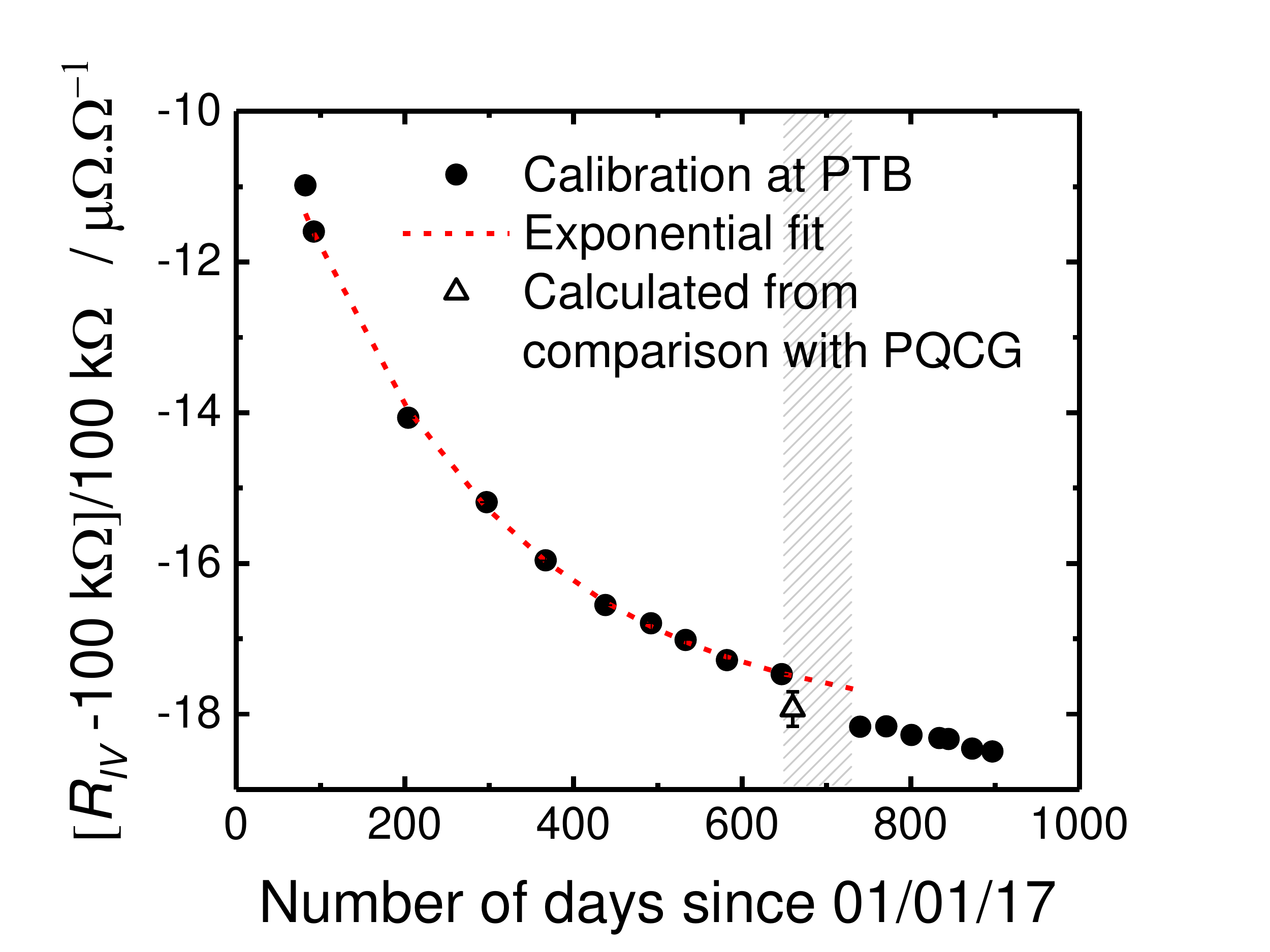}
  \caption{\small{Relative deviation of $R_{IV}$ from the nominal value 100 k$\Omega$ measured at PTB (black dots). The typical uncertainty of the output stage calibrations is 6$\times10^{-9}$ (much smaller than the size of the dots). The black triangle represents the equivalent quantity deduced from the measurements done at LNE, the associated error bar is the combined uncertainty without the contribution from the stability of the ULCA ($k=1$). All values refer to an ULCA-internal temperature of 23.0 $^{\circ}$C. The hatched area represents the traveling period out of PTB.}}
  \label{fig:ULCA_history}
\end{figure*}

 Figure\space\ref{fig:ULCA_history} shows the relative deviation of $R_{IV}$ from the 100-k$\Omega$ nominal value measured at PTB (black dots) as a function of time over a period of more than two years including the period of the comparison. Before transportation, the transresistance followed an exponential decay at PTB (red dashed line in figure\space\ref{fig:ULCA_history}). However, the first calibration after returning to PTB showed a deviation of $4.8\times10^{-7}$ from the predicted value, which might be due to mechanical and thermal shocks during transportation. Unfortunately, we cannot clearly assign the discontinuity to the transport to or from LNE. This impacts the estimation of the transresistance, and hence the value of $I_{\mathrm{ULCA}}$, at the time of the comparison. Here, to make the less assumptions possible, we have considered a linear interpolation at the date of the comparison, between the value of the last calibration before transportation and the first after return to PTB. The estimated uncertainty is calculated from a rectangular probability distribution of total width equal to the difference between these two values (7$\times10^{-7}$). The relative uncertainty on the estimation of $R_{IV}$ is therefore $2\times10^{-7}$. From the measurements done at LNE, the transresistance $R_{IV}$ can be estimated from $\Delta I/I= (R_{IV}-100 \mathrm{k}\Omega)/100 \mathrm{k}\Omega$. This value can be compared to the value obtained at PTB. It has been added in figure \space\ref{fig:ULCA_history}. The error bar reflects the total combined uncertainty without the contribution of the stability of the $R_{IV}$ and is therefore dominated by the contribution due to the calibration of the DVM. On a general point of view, in principle, the transresistance could also be directly calibrated by means of the CCC-based resistance bridge presented in \cite{Poirier2020} and traced back to the quantum Hall resistance at LNE.

\end{document}